\documentclass[twocolumn, apl, amsmath, amssymb, showpacs, superscriptaddress]{revtex4-2}
\usepackage{epsf}      
\usepackage{graphicx}
\usepackage{color}
\usepackage{soul}
\usepackage{gensymb}
\usepackage{sidecap}
\usepackage{amsmath}
\usepackage{mathtools}

\begin{document}

\title{The influence of Ga doping on  magnetic properties,  magnetocaloric effect, and electronic structure of  pseudo-binary GdZn$_{1-x}$Ga$_{x}$ (x = 0-0.1)}

\author{Anis Biswas}

\email{anis@ameslab.gov}

\affiliation{Ames National Laboratory, U.S. Department of Energy, Iowa State University, Ames, Iowa 50011, USA} 

\author{Ajay Kumar}

\affiliation{Ames National Laboratory, U.S. Department of Energy, Iowa State University, Ames, Iowa 50011, USA}

\author{Prashant Singh}
\affiliation{Ames National Laboratory, U.S. Department of Energy, Iowa State University, Ames, Iowa 50011, USA}

\author{Tyler Del Rose}
\affiliation{Ames National Laboratory, U.S. Department of Energy, Iowa State University, Ames, Iowa 50011, USA}

\author{Rajiv K. Chouhan }
\affiliation{Ames National Laboratory, U.S. Department of Energy, Iowa State University, Ames, Iowa 50011, USA}

\author{B. C. Margato}
\affiliation{Instituto de Física, Universidade do Estado do Rio de Janeiro - UERJ, Rua São Francisco Xavier 524, RJ 20550-013, Brazil}

\author{ B. P. Alho}
\affiliation{Instituto de Física, Universidade do Estado do Rio de Janeiro - UERJ, Rua São Francisco Xavier 524, RJ 20550-013, Brazil}

\author{E. P. N\'obrega}
\affiliation{Instituto de Física, Universidade do Estado do Rio de Janeiro - UERJ, Rua São Francisco Xavier 524, RJ 20550-013, Brazil}

\author{ P. J. von Ranke}
\affiliation{Instituto de Física, Universidade do Estado do Rio de Janeiro - UERJ, Rua São Francisco Xavier 524, RJ 20550-013, Brazil}

\author{P. O. Ribeiro}
\affiliation{Instituto de Física, Universidade do Estado do Rio de Janeiro - UERJ, Rua São Francisco Xavier 524, RJ 20550-013, Brazil}

\author{V. S. R. de Sousa}
\affiliation{Instituto de Física, Universidade do Estado do Rio de Janeiro - UERJ, Rua São Francisco Xavier 524, RJ 20550-013, Brazil}

\author{Yaroslav Mudryk}
\affiliation{Ames National Laboratory, U.S. Department of Energy, Iowa State University, Ames, Iowa 50011, USA}

\date{\today}

\begin{abstract}

We explore the impact of introducing  IIIA-group element Ga in place of  IIB-group element Zn in binary intermetallic GdZn on its magnetic and magnetocaloric properties, as well as explicate the modified electronic band structure of the compound. The magnetic transition temperature of the compound decreases with the increase of Ga concentration in GdZn$_{1-x}$Ga$_{x}$ (x = 0-0.1) while the crystal structure (CsCl-prototype) and lattice parameters remain unchanged. Our detailed analysis of magnetization and magnetocaloric data conclusively proves that long-ranged magnetic ordering exists in the sample, despite the magnetic interaction considerably weakening with the increase of Ga. The experimental data is rationalized using both  theoretical machine learning model and  first-principle density functional theory.The electronic band structure of GdZn is manifested with some unusual complex features which  gradually diminish with Ga doping and conventional sinusoidal feature of Ruderman-Kittel-Kasuya-
Yosida (RKKY)-type interactions also disappears. A mean-field theory model is developed and can successfully describe the overall magnetocaloric behavior of the GdZn$_{1-x}$Ga$_{x}$ series of samples.  

\end{abstract}

\maketitle

\section{\noindent ~Introduction}

The lanthanide-based intermetallic (LBI) compounds represent a unique class of materials with widely diverse and interesting magnetic properties, some owing their important features to concomitant magnetic and structural phase transitions \cite{Szytula1993, Gignoux1991, Savchenkov2023, Chinchure2002, Chakraborty2024, Biswas2024a, Yu2020, Biswas2020PRB, Alho2022, Ribeiro2022, Biswas2020Jalcom, Biswas2022, Ribeiro2024, DelRose2024, Petit2015}. Unique to most of these compounds (except La-based) is the presence of 4$f$ electrons, which adds a layer of complexity to their magnetism. It is generally accepted that the 4$f$ electronic states in LBIs are highly localized, and an indirect magnetic exchange occurs through Ruderman-Kittel-Kasuya-Yosida (RKKY)-type interactions, where $d$ electrons play a vital part \cite{Campbell1972, Haskel2007}. Lanthanides are usually trivalent when forming intermetallics \cite{Campbell1972}; however, there are a few examples of Eu-, Yb-, and Ce-based intermetallic compounds where lanthanide atoms are divalent \cite{Guillou2018, Guillou2020}. The long-range magnetic order in LBI is strongly dependent on the 4$f$-5$d$ hybridization and is also supported by itinerant $s$, $p$, and $d$ electrons. Those itinerant electronic states often come from constituent elements and can play a major role in mediating magnetic interactions \cite{Haskel2007}. As a consequence, the magnetic properties of LBI can also be greatly influenced by the presence of the non-lanthanide elements in such compounds even if those are non-magnetic. The comprehensive understanding of how does the presence of non-magnetic element modify the magnetic interactions in  LBI alloys can provide useful guidence to downselect suitable substitution element in order to design functional lanthanide based materials with tunable magnetic properties. \par

Among different lanthanides, the magnetic transition temperature of Gd is the highest, and several Gd-based alloys are well known to show many intriguing phenomena including large magnetocaloric effect (MCE), anomalous thermal expansion, spontaneous generation of thermoelectric power, magnetic memory, magnetic frustration, magnetic deflagration, exchange bias, large magnetoresistance, and topological Hall effect \cite{Pecharsky1997, Levin2001, Nazih2003, Chakraborty2022,  Velez2010, Morellon1998, DelRose2021, Kurumaji2019}. In particular, a group of binary alloys with the general formula GdZ (Z = elements of IB, IIB, and IIIA groups of the periodic table) show a rich variety of magnetic properties depending on the choice of Z and thus has been under intense scrutiny over the last few decades \cite{Kanematsu1969, Oppelt1972, Alfieri1966, Takei1979, Yashiro1976, Kobler1981}. In this context, one can recall that while  binary GdZ intermetallics with Z belonging to the group IB of the periodic table (e.g., GdCu, GdAg, and GdAu) show antiferromagnetic (AFM) transition \cite{Kissell1966, Sekizawa1966}, choosing Z from the  IIB group (e.g., Zn, Cd, and Hg) changes the magnetic ground state to ferromagnetic (FM) \cite{Petit2015, Rouchy1981, Petit2020}. Furthermore, experimental studies confirmed the AFM transitions in GdZ with Z = IIIA group elements such as Tl and In \cite{Sekizawa1966, Sekizawa1983}. A notable exception is found in the case of GdGa in which an FM ordering followed by a spin reorientation transition at low temperature \cite{Susilo2014} was revealed despite Ga being an element from IIIA. Such a strong variation in magnetic exchange interaction depending on the selection of Z clearly endorses the fact that the conduction electrons play a vital role in determining the types of magnetic ordering in those compounds. This is highly interesting because classic RKKY assumes strong changes in the sign of magnetic interactions depending on the interatomic distance. But in this case, the change in electronic concentration appears to be much more important than a change in lattice dimensions, e.g., between GdCu and GdAg.\par
 
The crystal structure of GdZ is a simple CsCl-type when Z is either from IB or IIB groups. However, the crystallography becomes more complex when Z belongs to IIIA. Namely, the CsCl structure is found for Z = Tl \cite{Baenziger1961} and In \cite{Delfino1983}, while crystal structures of other prototypes are evident in GdAl (DyAl-type) \cite{Buschow1965} and GdGa (CrB-type)\cite{Susilo2014}, even though both Al and Ga belong to IIIA. Despite considerable studies on GdZ type intermetallics with Z from either IB, IIB, or IIIA, little attention has been paid towards solid solutions between Gd and Z taking two elements in the Z site from two different groups of the periodic table, i.e., GdZ$^1_{1-x}$Z$^2_{x}$ (Z$^1$ and Z$^2$ belong to different groups).

 Recently, first-principles theoretical calculations were performed to describe magnetic ordering transitions of different GdZ$^1_{1-x}$Z$^2_{x}$ type pseudo binary alloys \cite{Petit2020}. Among those materials, the theoretical results for GdZn$_{1-x}$Ga$_x$ are quite interesting. According to the density functional theory (DFT) calculation, the magnetic transition temperature ($T_{\rm C}$) for GdZn$_{1-x}$Ga$_x$ is predicted to increase with $x$ for smaller doping concentrations ($x<0.1$) and to decrease almost linearly for higher $x$, indicating a significant difference in magnetic interactions between the lower and higher $x$ region of the series. However, $T_{\rm C}$'s of those alloys were considerably underestimated by DFT, and the predicted $x$-dependence of $T_{\rm C}$ for $x<0.1$ was not realized in experiments \cite{Petit2020}. This discrepancy between theory and experiment remains a knowledge gap that needs to be bridged for developing a predictive theory of magnetism in these and related compounds \cite{Petit2020}.  

There is clearly a need for a detailed study to expound the impact of Ga doping on magnetism of GdZn and its associated magnetocaloric effect (MCE), especially in the critical region of phase transition. Considering different electron count of Zn and Ga, such a study would be crucial to gain deep insights into the precise roles of elements in the Z site with different number of electrons in determining magnetic ordering of GdZ$^1_{1-x}$Z$^2_{x}$-type pseudo binary intermetallic alloys in general. This will also shed light on the underlying factors responsible for the apparent difference between theoretical and experimental $T_{\rm C}$'s of the pseudo binary series of GdZn$_{1-x}$Ga$_x$ compounds as presented in the earlier work \cite{Petit2020}. Furthermore, a recent electronic structure calculation for GdZn revealed an unusual electronic band structure of the compound manifesting nodal lines at the Fermi level, making the material a fertile platform for exploring the interplay between topology and magnetism \cite{Zhao2021}. Thus, a relevant question may arise regarding how that band structure is modified due to Ga replacing Zn in the compound. \par

Here we present a comprehensive study combining both theory and experiment to understand the detailed magnetic along with magnetocaloric properties, and electronic structure of GdZn$_{1-x}$Ga$_x$ with special emphasis on the explication of the possibility of anomalous doping dependence of magnetic transitions of the compounds.  A recently developed machine learning-based predictive model was used to evaluate $T_{\rm C}$'s of substituted materials. A theoretical model based on mean-field theory was formulated to describe the experimentally obtained MCE of the alloys. Additionally, we address an important fundamental question regarding why predictive theoretical calculation is challenging for GdZn$_{1-x}$Ga$_x$ compounds.

\section{\noindent ~Methodology}

\subsection{\noindent ~Experiment}

We prepared a series of alloys with compositions GdZn$_{1-x}$Ga$_x$ (where $x = 0$, 0.05, and 0.1) weighing 5 g each by melting the stoichiometric amounts of constituent elements using a high-frequency induction furnace. Among the constituent elements, Zn and Ga metals were purchased from Alfa Aesar with a purity of 99.99 at.\%, whereas Gd (99.8 at.\% with respect to all elements, including O, C, and N) was secured from the Materials Preparation Center, Ames National Laboratory, USA \cite{AmesLab}. The melting was done inside a tantalum crucible sealed in a partial (0.8 atm) helium atmosphere. Prior to sealing, the empty tantalum crucibles were heated for 2 hours at 1800 $^o$C for purification from different absorbed gases. Each of the sealed crucibles containing the constituent elements was placed inside the induction furnace at $\sim 1500 ^o$C for 2 hours with intermediate flipping of the crucibles for complete melting and homogeneous mixing. After melting, the prepared samples were removed from the crucibles and heat-treated at 700 $^o$C for 5 days inside He-filled quartz tubes.\par

For structural characterization, room temperature powder X-ray diffraction (PXRD) was performed on a modified Rigaku TTRAX system with Mo K$\alpha$ radiation \cite{Holm2004}. Rietveld refinement of PXRD data was performed using the GSAS-II software \cite{Toby2013}. For scanning electron micrographs and elemental analysis, we utilized an FEI Teneo Scanning Electron Microscope (SEM) equipped with an Oxford Instruments Aztec Energy Dispersive Spectroscopy (EDS) system.

The X-ray Photoelectron Spectroscopy (XPS) measurements were performed using a Kratos Amicus/ESCA 3400 instrument. We perform a very mild Ar sputtering of 15 sec to remove the surface contamination without significantly disturbing the surface stoichiometry of the samples.  The sample was irradiated with 240 W unmonochromated Al K$\alpha$ x-rays, and photoelectrons emitted at 0° from the surface normal were energy analyzed using a DuPont type analyzer. The pass energy was set at 150 eV. All spectra were energy calibrated with measured C 1$s$ peak position at 284.8 eV after subtracting the inelastic Tougaard background. The sum of Voight functions have been used to fit the core-level spectra. 

The magnetic properties of the samples were studied using a superconducting quantum interference device (SQUID) magnetometer (MPMS XL-7 by Quantum Design). We studied the temperature dependence of DC magnetization of the samples, $M(T)$, in the temperature range 5-395 K at 500 Oe and 20 kOe. The minimum of the first derivative of $M(T)$ with respect to temperature, i.e., the temperature corresponding to the fastest change in magnetization, is roughly defined as $T_{\rm C}$. However, the broadness of the transition in the case of a second-order magnetic transition contributes significant error in the determination of $T_{\rm C}$ using this method. Hence, we estimated $T_{\rm C}$ and different critical exponents associated with the transition from isothermal $M(H)$ curves recorded at different temperatures around the magnetic transition using modified Arrott-Noakes plots following the Kouvel-Fisher method \cite{Kouvel1964}. The $M(H)$ curves were recorded in the first quadrant (0 to 70 kOe) for all the samples at different temperatures across the transition. The magnetic field-induced entropy changes of the samples, $\Delta S$, were determined from those $M(H)$ data using Maxwell’s relations to evaluate the magnetocaloric behavior of the samples \cite{NevesBez2018}. To examine the magnetic ground state, full $M(H)$ loops of the samples were recorded  in the magnetic field ranging from -50 kOe to +50 kOe at 5 K.

\subsection{\noindent ~Theory}
A theoretical model was formulated based on mean-field theory (MFT) \cite{Biswas2020Jalcom, Alho2022, Ribeiro2022} to quantify magnetocaloric properties for the GdZn$_{1-x}$Ga$_x$ ($x = 0 - 0.1$) samples as the analysis of their magnetization  data  can be successfully  described by MFT (discussed latter). 

We  performed electron-structure calculations for the compounds using DFT  as implemented within Vienna Ab initio Simulation Package (VASP)\cite{Hafner2008} and also using a full-potential linearized augmented plane wave (FP-LAPW) method as implemented in the code WIEN2k\cite{Blaha2020}.
In the VASP based DFT, the valence interaction among electrons were described by a projector augmented-wave method \cite{Kresse1, Monkhorst1}  with a plane-waves energy cutoff of 520 eV. For full relaxation of compounds, we set very strict convergence criteria of total energy and force convergence, i.e., 10$^{-8}$ eV/cell and 10$^{-6 }$eV/\AA. In (semi)local functionals, such as GGA, the $f$-electrons are always delocalized due to their large self-interaction error\cite{PS1,PS2}, so we employ the Perdew-Burke-Ernzerhof (PBE) exchange-correlation functional in the generalized gradient approximation (GGA)\cite{Perdew1996}.   To enforce the localization of the $f$-electrons, we perform PBE+U calculations [with a Hubbard U (Gd=6.6 eV; Zn=7.5 eV; Ga=6 eV; J=0.9 eV) introduced in a screened Hartree-Fock manner\cite{Dudarev1}. We used 7 × 7 × 7 (11 × 11 × 5) and Monkhorst-Pack k-mesh for Brillouin zone sampling of Gd-Zn-Ga compounds during structural optimization (electronic-relaxation) \cite{Monkhorst1}. The partially order structures of GdZn$_{1-x}$Ga$_{x}$ (x=0, 0.05, and 0.1) for electronic-structure calculations were generated using Alloy Theoretic Automated Toolkit (ATAT)\cite{VanDeWalle2002}.

\begin{figure*}  
\centering
\includegraphics[width=1\textwidth]{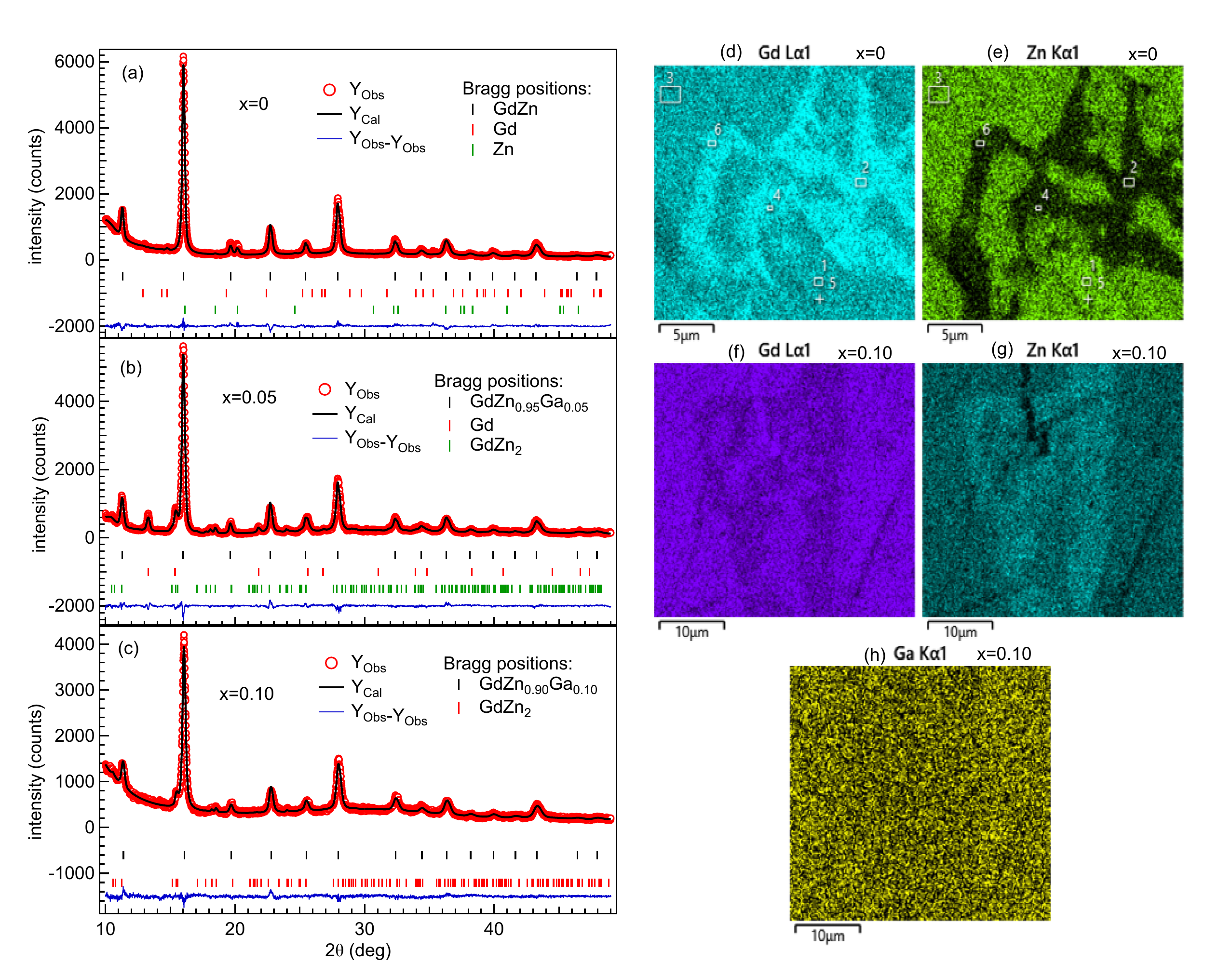}
\caption {(a--c) Room temperature  powder x-ray diffraction data of GdZn$_{1-x}$Ga$_x$ for $x$ = 0, 0.05, and 0.10 samples, respectively, using Mo K$\alpha$ ($\lambda \approx$0.71 \AA) radition. The elemental mapping at the selected areas on (d, e) $x =$ 0 and (f-h) $x =$ 0.10 samples using energy dispersive
spectroscopy. } 
\label{Fig1_XRD}
\end{figure*}

On the other hand, in FP-LAPW calculations, we also adopted the spin-polarized framework within the GGA as proposed by Perdew et al \cite{Perdew1996}. The core states were treated fully relativistically, whereas for the valence states, a scalar relativistic approach was used. Valence electronic wavefunctions inside the Muffin-tin sphere were expanded up to lmax = 10. The atomic radii for Gd, Zn, and Ga are set as 2.5 bohr with a force minimization of 2, respectively to achieve nearly touching spheres avoiding the charge leakage with a force minimization of 0.01 Ry/a.u.  The plane wave cut-off parameters were setteled by R$_{MT}$k$_{max}$ = 7 (where k$_{max}$ is the largest wave vector of the basis set) and Gmax = 12 a.u.$^{-1}$ for Fourier expansion of potential in the interstitial region with the energy separation of -6.0 Ry between valance and core states. To ensure high accuracy in calculations a dense k-mesh of 18 x 18 x 18 is performed. The k-mesh is further adjusted according to the supercell size selected for the Ga doping. For the accuracy of the self-consistent calculations the convergence criteria of charges and energies were set to be 10$^{-5}$ Ry and 10-6 Ry, respectively. Proper treatment of strongly localized 4$f$ electron are done by including the Hubbard U approach, where the optimized onsite electron correlation parameters U$_{eff}$ = U – J = 6 eV is set as used in earlier study\cite{Zhao2021}. The Pm-3m CsCl-type structure of GdZn with Gd, and Zn atoms occupy the sites 1a (0, 0, 0), and 1b (0.5, 0.5, 0.5) in the unit cell. The primitive structure is relaxed and the optimized lattice parameters a = 3.608 Å found for GdZn, that is close to the experimental value 3.604 \AA. Therefore, we used the experimental lattice parameters for the calculation.

Finally, to assess $T_{\rm C}$’ s of the samples, we used random forest (RF)  machine-learning (ML) model trained on experimentally known ferromagnetic and ferrimagnetic compounds \cite{Singh2023,Nelson2019}. The details of the ML model training, test, and validation are provided elsewhere \cite{Singh2023}. The regressor performence was quantified using the coefficient of determonation (R$^{2}$).  The crossvalidation of our model was done with R$^{2}$ = 0.91, demonstrating high confidence in estimations of $T_{\rm C}$’ s. 

\section{\noindent ~Results and discussion}

\subsection{Phase composition and crystal structure}

\begin{table*}
    \centering
    \caption{Rietveld refinement results of the studied GdZn$_{1-x}$Ga$_x$. The last column shows the actual Ga content ($x$) determined by EDS.}
    \begin{tabular}{p{2cm}p{2.5cm}p{2cm}p{2cm}p{2cm}p{2cm}p{2cm}p{2cm}}
        \hline
        Sample & Phase & Space Group & Wt.\% & $a$ (\AA) & $b$ (\AA) & $c$ (\AA) & Ga($x$): EDS \\
        \hline
        $x = 0$ & GdZn & Pm-3m & 96.4 & 3.6045(3) & -- & -- & - \\
         & Zn & P63/mmc & 2.6 & 2.555(1) & -- & 5.062(2) & - \\
         & Gd & P63/mmc & 1.0 & 3.65(1) & -- & 5.70(2) & - \\
        \hline
        $x = 0.05$ & GdZn$_{0.95}$Ga$_{0.05}$ & Pm-3m & 79.3 & 3.5988(3) & -- & -- & 0.053 \\
         & Gd & Fm-3m & 9.8 & 5.229(1) & -- & -- & - \\
         & Gd(Zn/Ga)$_2$ & Imma & 10.9 & 4.504(1) & 7.210(3) & 7.594(2) & 0.040 \\
        \hline
        $x = 0.10$ & GdZn$_{0.90}$Ga$_{0.10}$ & Pm-3m & 87.0 & 3.6035(4) & -- & -- & 0.102 \\
         & Gd(Zn/Ga)$_2$ & Imma & 13.0 & 4.5021(3) & 7.279(6) & 7.620(4) & 0.033 \\
        \hline
    \end{tabular}
    \label{T_XRD}
\end{table*}

\begin{figure*} 
\includegraphics[width=7in]{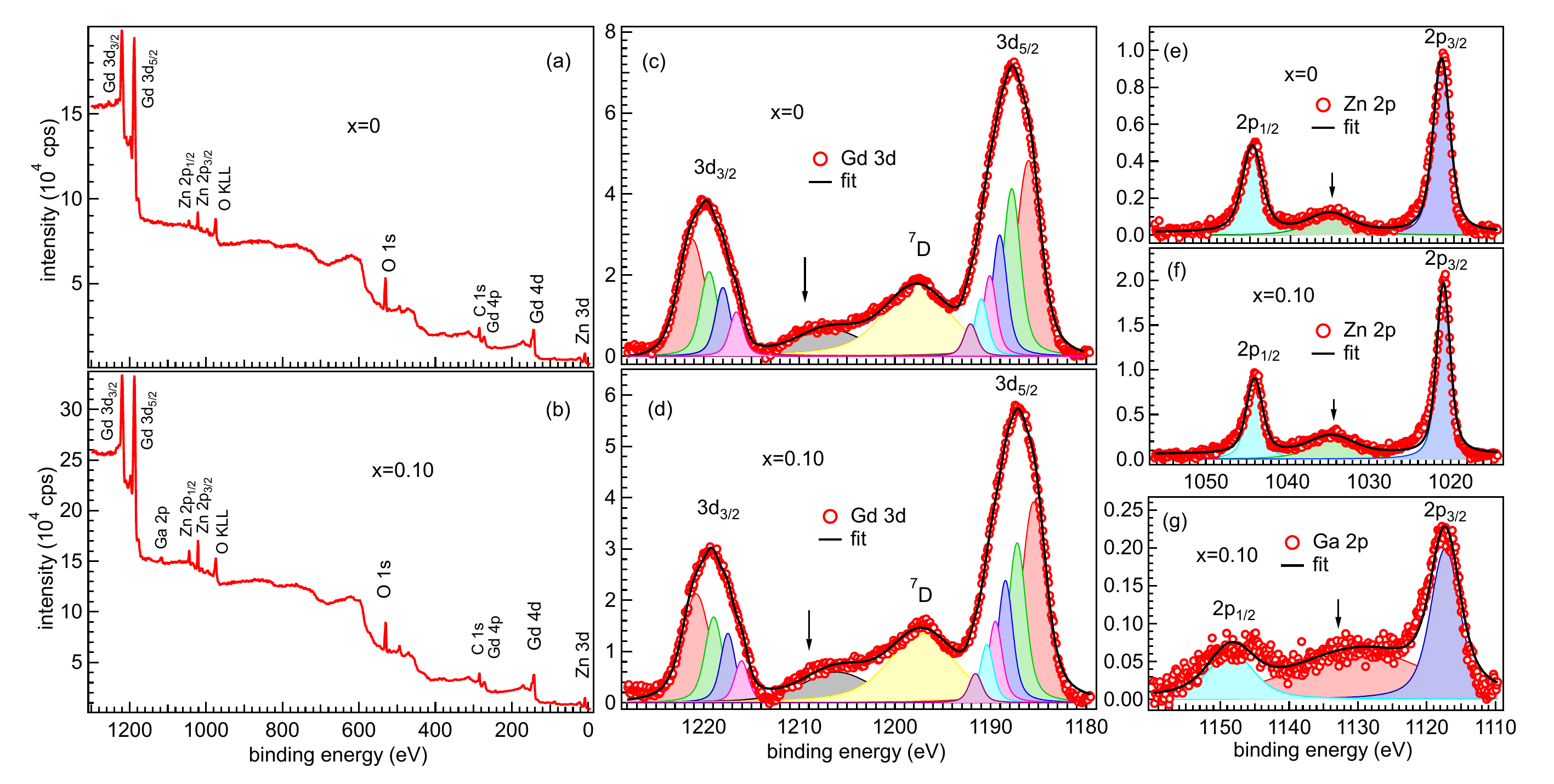}
\caption {(a, b) The XPS survey spectra, (c, d) Gd 3$d$ core-level, and (e, f) Zn 2$p$ core-level spectra of the GdZn$_{1-x}$Ga$_x$ for $x=0$ and 0.10, respectively. (g) Ga 2$p$ core-level spectra of $x =$ 0.10 sample.} 
\label{Fig2_XPS}
\end{figure*}

The PXRD data at room temperature [Figs. \ref{Fig1_XRD}(a--c)] reveal that the main phase crystallizes in the CsCl-type crystal structure (cubic) with other impurity phases present in all samples (see Table \ref{T_XRD}). In GdZn, a small amount of unreacted Gd and Zn (both in hexagonal structure with space group P6$_3$/mmc) is detected, whereas a common GdZn$_2$ impurity phase was observed for both Ga-doped samples, and its relative weight percentage is slightly larger in GdZn$_{0.90}$Ga$_{0.10}$ compared to GdZn$_{0.95}$Ga$_{0.05}$ (Table \ref{T_XRD}). Note that from the XRD study, the presence of a small amount of Ga at the Zn site in this secondary phase can not be probed due to the close x-ray scattering factors of Ga and Zn atoms. The presence of a face-centered cubic (fcc) type Gd phase, which is different from both the regular hexagonal (hcp) structure and the high-temperature body-centered cubic (bcc) polymorph, is also realized for GdZn$_{0.95}$Ga$_{0.05}$. To the best of our knowledge, so far the fcc Gd has been reported only for thin films, but the lattice parameter $a$ for those reported cases is considerably smaller than in our case \cite{Curzon1973, Liu2020}. Therefore, it is possible that this phase is stabilized by interstitial elements, for example, by mild hydrolysis during sample preparation. This would also explain why this Gd-rich phase does not show up in magnetic data (shown later) as would be expected for hcp Gd.

We also performed SEM-EDS studies on the $x =$ 0 and 0.10 samples, as shown in Figs. \ref{Fig1_XRD}(d-h). According to the EDS results, the overall compositions of the main phase practically retained the desired stoichiometric ratios of the constituent elements with good homogeneity.  Notably, EDS results confirmed that GdZn ($x $= 0) sample has a non-uniform Gd impurity distribution, forming within binary GdZn phase [see elemental mapping in Figs. \ref{Fig1_XRD}(d, e)]. On the other hand, the EDS spectra recorded at the different positions on the $x =$ 0.10 sample also show a segregation of Zn rich secondary phase within targeted GdZn$_{0.90}$Ga$_{0.10}$ main matrix. The elemental color mapping shown in Figs. \ref{Fig1_XRD}(f, h) clearly demonstrate the presence of a Zn-rich phase in the sample, with a composition Gd(Zn$_{0.91}$Ga$_{0.03}$)$_2$, which is close to the GdZn$_2$ composition, as also evident from the refinement of the XRD data. The presence of unreacted Zn in XRD and Gd in the binary sample is puzzling given the high temperature of the synthesis. We hypothesized that a large volume of Zn remained gaseous during the reaction and did not partake in the reaction, precipitating upon cooling. However, the addition of extra Zn leads to the formation of the GdZn$_2$-based phase, which is also quite stable. No extra Zn was added in our experiment, but the inhomogeneous distribution of Zn across the crucible likely led to the multiphase nature of the sample. We conclude that the formation of 100\% pure GdZn$_{1-x}$Ga$_x$ phase may be difficult to achieve using our technique. However, encouraged by the stoichiometric composition of the main phase, and the fact that the presence of the impurity phase should not significantly impact our magnetic study, as discussed below, we decided to continue our study of these samples. The CsCl-type crystal structure of the parent compound (GdZn) is preserved despite the substitution of Zn by Ga in GdZn$_{1-x}$Ga$_x$ in small quantities ($x = 0-0.1$), even though GdGa is known to form with the CrB-type crystal structure \cite{Petit2020}. It appears that the substitution of Ga hardly makes any noticeable change in lattice parameters or lattice volume within the experimental error, despite causing a significant change in T$_{\rm C}$ (discussed below). The minimal change in the lattice parameters is not surprising considering the fact that the difference in atomic radii between Ga and Zn is  small \cite{Teatum1968}, and only minor Ga substitutions are examined here.

\begin{figure*} 
\includegraphics[width=5in]{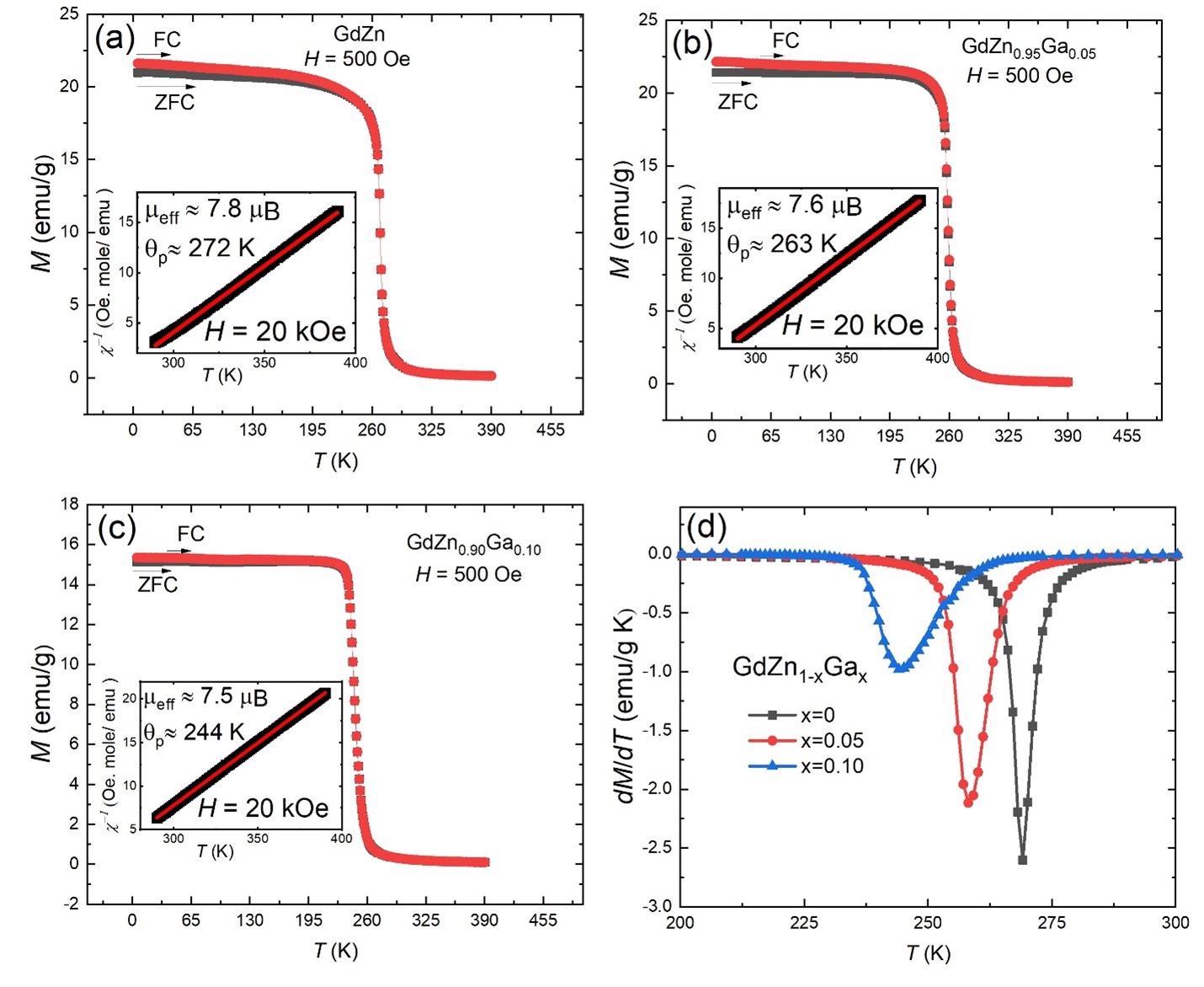}
\caption {The temperature dependencies of ZFC and FC magnetization curves of GdZn$_{1-x}$Ga$_x$ for (a) $x=0$, (b) $x=0.05$, and (c) $x=0.1$. The measurements were conducted at $H = 500$ Oe. The Curie-Weiss fittings of $\chi_{dc}^{-1}$ vs. $T$ in the paramagnetic temperature range (data recorded in the presence of a 20 kOe magnetic field) for the samples are shown as insets in (a), (b), and (c). The temperature dependence of $dM/dT$ for the samples is shown in (d).} 
\label{Fig3_MT}
\end{figure*}

\begin{figure} 
\includegraphics[width=3.4in, height=6in]{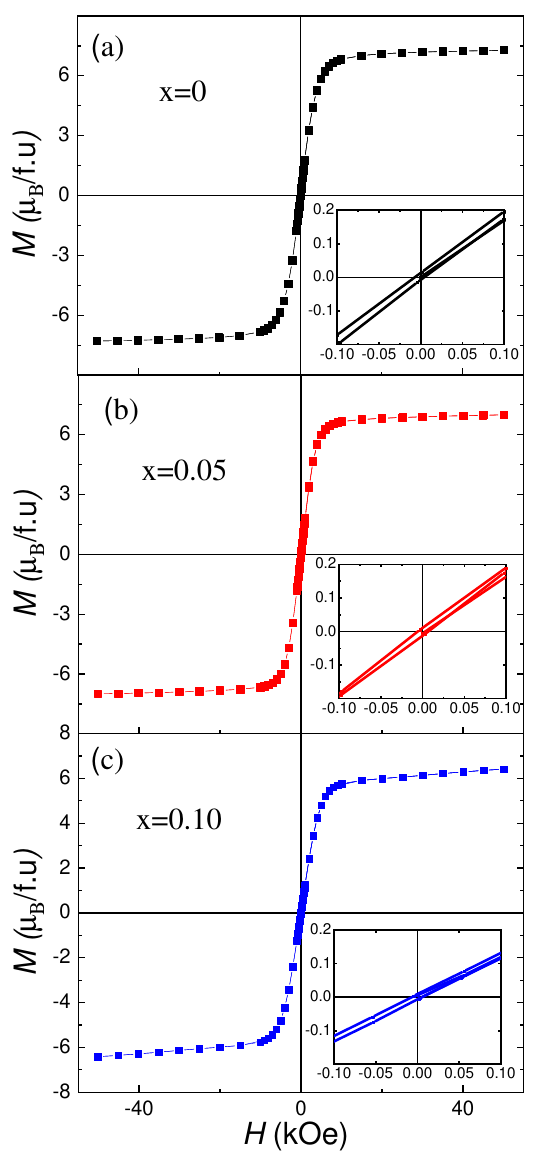}
\caption {M vs. H curves recorded at 5 K for GdZn$_{1-x}$Ga$_x$ for (a) $x=0$, (b) $x=0.05$, and (c) $x=0.1$. The expanded views of the curves in the low magnetic field region are shown in insets.} 
\label{Fig4_MH}
\end{figure}

\subsection{X-ray photoemission spectroscopy}

The XPS measurements were performed on the $x = 0$ and 0.10 samples to understand the electronic states of the constituent elements, as shown in Figs. \ref{Fig2_XPS}(a--g). Figs. \ref{Fig2_XPS}(a, b) display the survey spectra of the $x = 0$ and 0.10 samples, respectively, where all the prominent features are assigned to the binding energies of the constituent elements, ruling out the possibility of any elemental impurities in the samples, except for the presence of a sharp oxygen 1$s$ peak on the surface of both samples. The Gd 3$d$ spectra are shown in Figs. \ref{Fig2_XPS}(c) and (d) for the $x = 0$ and 0.10 samples, respectively. Two broad features centered around $\sim$1188 and 1220~eV are attributed to the spin-orbit split components 3$d_{5/2}$ and 3$d_{3/2}$ of Gd, respectively. Further inspection demonstrates the asymmetry in these features on both sides of the maxima, which, along with their significant broadening, indicates the contribution of more than one component to the peaks.

The Gd 3$d$ core-level hard X-ray photoemission spectroscopy measurements performed on Gd-based intermetallics demonstrate the fine splitting of the 3$d_{5/2}$ and 3$d_{3/2}$ levels due to the intermediate coupling scheme of the Gd$^{3+}$ ions, displaying features of both L-S and j-J coupling \cite{Elmers_PRB_13, Chuan_PRB_20, Nguyen_PRB_22}. Following the same splitting scheme, we fit the 3$d_{5/2}$ and 3$d_{3/2}$ distributions with six and four components, respectively. A $j-J$ coupling of the J = 7/2 state of the 4$f^7$ configuration with the $j = 5/2$ core hole state of 3$d_{5/2}$ gives $J' = 6, 5, 4, 3, 2, 1$ multiplets. This is analogous to the L-S coupling of the $^9$D state of 3$d_{5/2}$ with the $^8$S state of Gd$^{3+}$ having a 4$f^7$ configuration. The lowest binding energy of $J' = 6$ is attributed to the parallel alignment of the spin and magnetic moments of the 3$d$ states with the spin moment of the 4$f$ shell \cite{Elmers_PRB_13, Chuan_PRB_20, Nguyen_PRB_22}. Furthermore, the coupling of the J = 7/2 state with the $j = 3/2$ core hole state of 3$d_{3/2}$ gives $J' = 5, 4, 3, 2$ multiplets. For 3$d_{3/2}$, $J' = 2$ has the lowest binding energy due to the parallel alignment of the 4$f$ and 3$d$ spin moments, which align antiparallel to the 3$d$ orbital moment.

We also observe a broad peak around 1197 eV, which is attributed to the $^7$D state arising from the L-S coupling between the core hole $^2$D state (3$d^9$) and the $^8$S (4$f^7$) state \cite{Elmers_PRB_13, Chuan_PRB_20, Nguyen_PRB_22}. Both $x = 0$ and 0.10 samples samples display broad plasmon features around 1207 eV, as indicated by arrows in Figs. \ref{Fig2_XPS}(c, d), respectively. It should be noted that the resolution of the data in this case limits the unique identification of all the multiplets of the 3$d_{5/2}$ and 3$d_{3/2}$ states, but the clear evidence of the $^7$D states and plasmon features, analogous to binary GdNi \cite{Chuan_PRB_20} and GdFe \cite{Elmers_PRB_13}, indicates the presence of trivalent gadolinium with an intermediate coupling scheme, exhibiting signatures of both L-S and j-J couplings in both samples.

The Zn 2$p$ core-level spectra for $x = 0$ and $x = 0.10$ are shown in Figs. \ref{Fig2_XPS}(e) and (f), respectively. Both samples exhibit the spin-orbit split components: 2$p_{3/2}$ at $\sim$1021.5 eV and 2$p_{1/2}$ at 1044.7 eV, with an energy difference of 23.2 eV. These values correspond to the metallic state of zinc (Zn$^0$) in the samples \cite{Samanta_NA_20, Feliu_AM_03, Biesinger_ASS_10}. Both samples also show broad plasmon loss features around 13.7 eV above the 2$p_{3/2}$, as indicated by arrows in Figs. \ref{Fig2_XPS}(e) and (f) \cite{Nayak_PRB_15}. Interestingly, the Zn 2$p$ spectra of the $x = 0$ sample exhibit greater broadening compared to the $x = 0.10$, whereas the Gd 3$d$ spectra display similar broadening in both samples [see Fig. 1(a, b) of \cite{SI}]. This indicates that Ga doping only alters the local chemical environment and, hence, the electronic states of the Zn-site of the compounds. This may result in a change in the RKKY interactions mediated by these states, even though the electronic state of the magnetic Gd$^{3+}$  ions remains the same.

The Ga 2$p$ spectra for the $x = 0.10$ sample show 2$p_{3/2}$ and 2$p_{1/2}$ components at 1117.3 eV and 1148.5 eV, respectively, along with a broad plasmon feature around 1129.5 eV, as indicated by an arrow in Fig. \ref{Fig2_XPS}(g). These binding energy values lie between the metallic (Ga$^0$) and oxide (Ga$^{3+}$) states of Ga atoms \cite{Chang_APL_12, Lee_NC_23}. However, the low concentration of Ga in the sample limits the precise identification of the two states. Therefore, we synthesized and performed XPS measurements on $x = 0.20$ sample, where the two peaks can be clearly deconvoluted (see Fig. 2 of \cite{SI}), indicating the possibility of surface oxidation of the Ga atoms.

\subsection{Magnetic properties}

The temperature dependent magnetization have been performed in zero field cooled (ZFC) and field cooled (FC) modes at 500~Oe for all the samples, as shown in Figs. \ref{Fig3_MT}(a--c) for $x =$0--0.1, respectively. In the ZFC mode, the sample was cooled down to 5 K in zero magnetic field, and then $M(T)$ data were acquired during the warming cycle in the presence of $H = 500$ Oe, whereas in the FC protocol the magnetization data were recorded at 500 Oe during the warming cycle after cooling the sample to 5 K in the presence of the same magnetic field. All the samples display a  ferromagnetic to a paramagnetic transition with small low-temperature irreversibility between FC and ZFC magnetization curves [ Figs. \ref{Fig3_MT}(a--c)]. At first approximation, we consider the temperatures corresponding to the minima of d$M$/d$T$ as $T_{\rm C}$ for the samples, as presented in Fig. \ref{Fig3_MT}(d) and Table \ref{T_MT}, which monotonically decreases with increase in the Ga concentration. The temperature dependence of dc magnetic susceptibility follows the Curie-Weiss law above the magnetic transition, as manifested by linear $\chi_{\text{dc}}^{-1}$ vs. $T$ curves [insets of Figs. \ref{Fig3_MT}(a--c)]. The obtained effective paramagnetic moments ($\mu_{\text{eff}}$) are in agreement with the respective theoretical $g\sqrt{J(J+1)}$ values for the samples, and the Weiss temperatures, $\theta_P$, are also close to the respective $T_{\rm C}$ values (Table \ref{T_MT}). Since both GdZn$_2$ and GdGa$_2$ order magnetically at 68 and 12 K, respectively \cite{Debray1970, Tsai1979}, much lower temperatures than $T_{\rm C}$ of the main phase , we can safely rule out the presence of any additional ordered magnetic phase of the impurity Gd(Zn$_{0.91}$Ga$_{0.03}$)$_2$  phase near $T_{\rm C}$ for the Ga-doped samples. \par

\begin{table}
    \centering
    \caption{Different magnetic parameters obtained from M(T) and M(H) data of GdZn$_{1-x}$Ga$_x$. The first column reports magnetic ordering temperature determined as a minimum of dM/dT. The effective paramagnetic moment, $\mu_{\text{eff}}$, the saturation moment, $M_S$, and the paramagnetic Weiss temperature, $\theta_P$, are shown for all samples.}
    \begin{tabular} {p{1.5cm}p{2cm}p{1.5cm}p{1.5cm}p{1.5cm} }
        \hline
        Sample & (dM/dT)$_{\rm min}$ (K) & $\mu_{\text{eff}}$ ($\mu_B$) & $M_S$ ($\mu_B$) & $\theta_P$ (K) \\
        \hline
        $x=0$ & 269 & 7.8 & 7.36 & 272 \\
        $x=0.05$ & 258 & 7.6 & 7.06 & 263 \\
        $x=0.1$ & 244 & 7.5 & 6.4 & 244 \\
        \hline
    \end{tabular}
    \label{T_MT}
\end{table}

The $M(H)$ curves for the GdZn$_{1-x}$Ga$_x$ samples at 5 K are shown in Figs. \ref{Fig4_MH}(a--c) for $x =$0--0.1, respectively, and confirm the ferromagnetic nature of the samples with saturation magnetization ($M_S$) close to their theoretical $gJ = 7 \mu_B / \text{Gd}$ values (Table \ref{T_MT}). We calculated $M_S$ by taking the molar mass of nominal compositions of each sample. As the impurity phases exist in the sample, the actual $M_S$ value for the samples would be slightly different if the masses of the impurity phases are taken into account in the calculation. None of the samples show any noticible magnetic hysteresis [insets of Figs. \ref{Fig4_MH}(a--c)], as expected for Gd-based (L=0) intermetallics due to the absence of magnetocrystalline anisotropy.

\subsection{Critical behavior}

In order to further unravel the magnetic interactions in these samples in the critical region around the ferromagnetic (FM) to paramagnetic (PM) transition, we extracted the critical exponents, $\beta$, $\gamma$, and $\delta$, associated with the transition. The values of these exponents can be used to analyze the nature of magnetic interactions and in the vicinity of $T_{\rm C}$ they can be expressed via the following equations of state \cite{Kouvel1964, Kouvel1968}:

\begin{figure*} 
\includegraphics[width=7in]{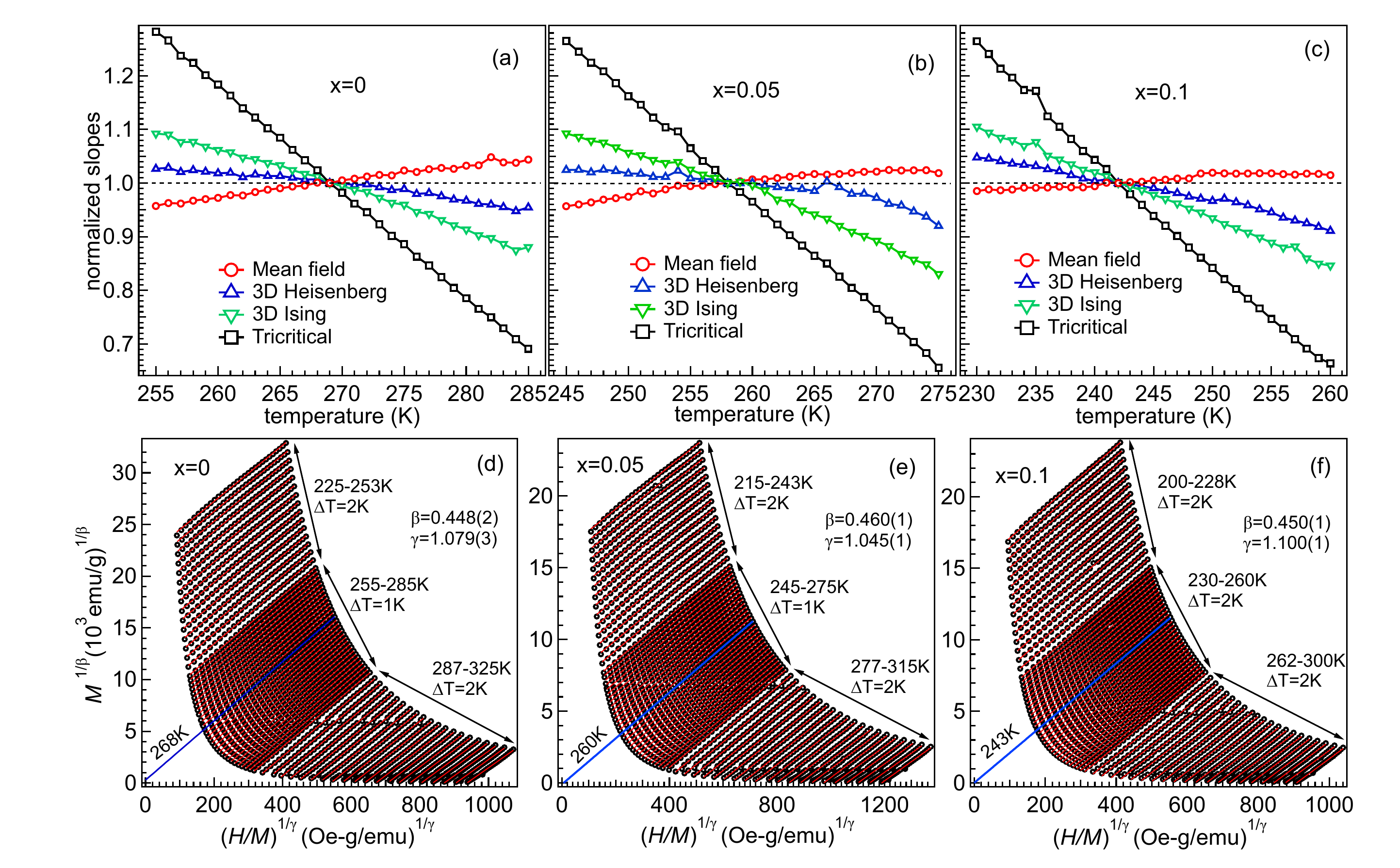}
\caption {(a--c) The normalized slopes (NS's) for different universality classes in the critical region for GdZn$_{1-x}$Ga$_x$ with $x= 0$, $0.05$, and $0.1$ respectively. (d-f): Modified Arrott plots using the correct critical exponents for the samples, where black circles, red, and blue lines represent the experimental data points, linear fit for $H > 10$ kOe, and the extrapolation of the linear fit at $T_C$ down to $H=0$.} 
\label{Fig5_Critical1}
\end{figure*}

\begin{eqnarray}
 M_{SP}(0, T) = M_0 (-\epsilon)^{\beta} \quad \text{for } \epsilon < 0, \quad T < T_C \\
 \chi_0^{-1}(0, T) = \left(\frac{h_0}{M_0}\right) \epsilon^{\gamma} \quad \text{for } \epsilon > 0, \quad T > T_C \\
 M(H, T_C) = D H^{1/\delta} \quad \text{for } \epsilon = 0, \quad T = T_C 
\end{eqnarray}
Here, $\epsilon = (T - T_C)/T_C$ is the reduced temperature, $M_{SP}$ is the spontaneous magnetization below $T_C$, and $\chi_0^{-1}$ is the initial inverse susceptibility above $T_{\rm C}$, whereas $M_0$, $h_0/M_0$, and $D$ are the critical amplitudes. Therefore, the magnetization isotherms were recorded from 0 to 70 kOe at different temperatures across the FM to PM transition for all the samples. The field was reduced from 70 kOe to 500 Oe in the linear mode and then from 500 Oe to 0 Oe in the oscillatory mode before moving to the next temperature to avoid any remanence in the samples. We first constructed the modified Arrott plots (MAPs) $[M^{1/\beta} Vs. (H/M)^{1/\gamma}]$ based on the Arrott-Noakes equation of state \cite{Arrott1967}: 

\begin{eqnarray}
\left(\frac{H}{M}\right)^{1/\gamma} = a \epsilon + b M^{1/\beta} 
\end{eqnarray}

where $a$ and $b$ are the temperature-dependent parameters. The MAPs for the mean field, 3D Heisenberg, 3D Ising, and Tricritical models, using the critical exponents listed in Table \ref{T_critical}, are shown in Figs. 1--3 of \cite{SI} for $x = 0$, $0.05$, and $0.1$ samples. The positive slopes of the curves confirm the second-order nature of the transition in all the samples. We fitted the MAPs using the linear equation (as shown by red lines in Figs. 1--3 of \cite{SI}) in their high magnetic field region ($H >$10 kOe). It appears that none of the universality classes display linear behavior in the entire temperature range, indicating the possibility of deviation from all the standard universality classes for these samples. We plotted the temperature-dependent normalized slope (NS) $= S(T)/S(T_{\rm C})$ for all the samples, where the slope of MAPs for any given temperature, $S(T)$, was normalized to its slope at $T_{\rm C}$. Figs. \ref{Fig5_Critical1}(a--c) show the NS for the $x = 0-0.1$ samples, respectively, using the $T_{\rm C}$ values from Table \ref{T_MT}. It is clearly evident that all the samples show the least deviation from unity for the mean field and 3D Heisenberg models as we move away from the critical point $T_{\rm C}$, indicating that either of those two models would successfully describe the magnetic interactions in the samples.

\begin{figure*} 
\includegraphics[width=6in]{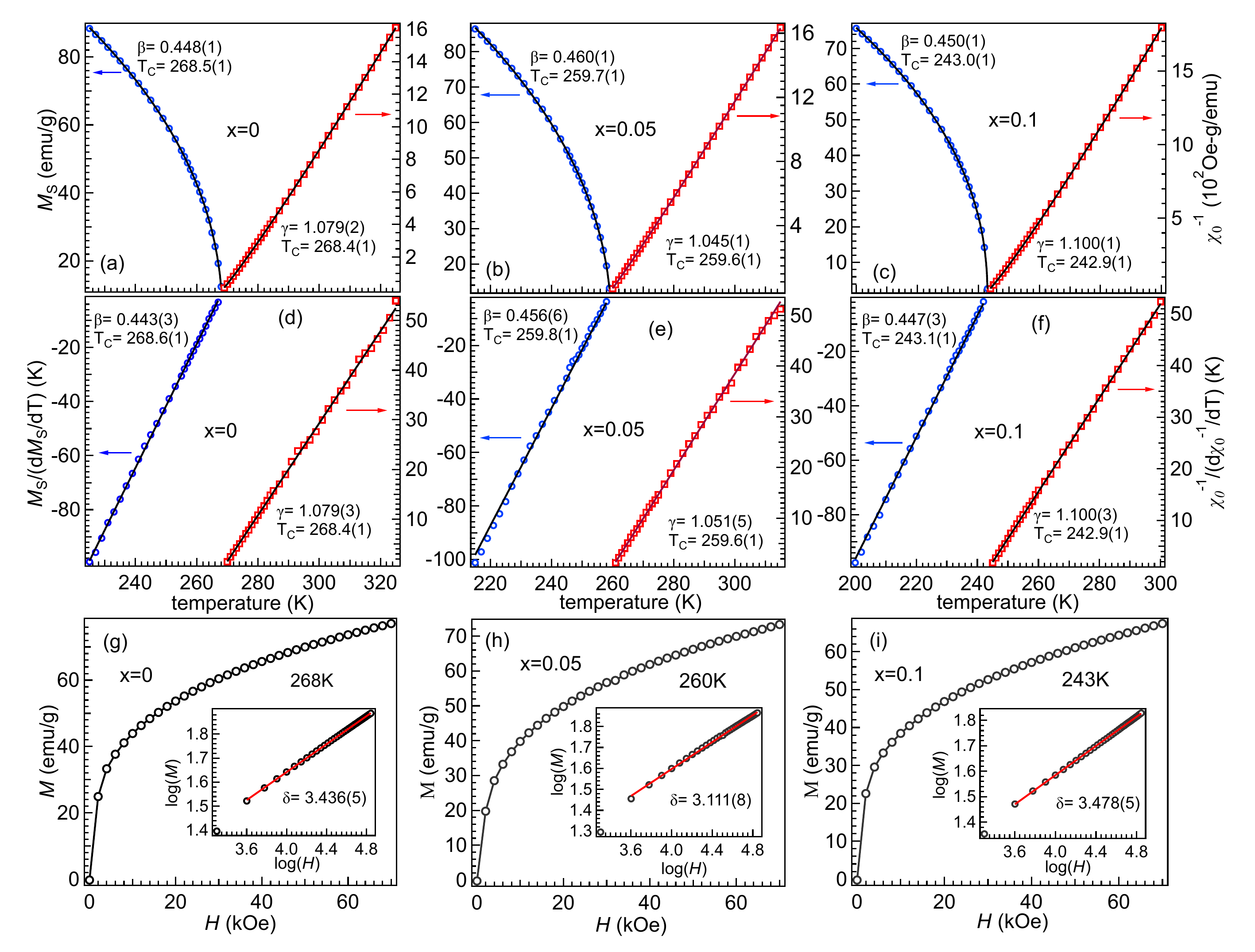}
\caption {(a-c) Temperature-dependent spontaneous magnetization $M_{SP}$ (on left axis) and inverse initial susceptibility $\chi_0^{-1}$ (on right axis), where black solid curves represent the best fit using Eqs. 1 and 2, respectively. (d-f) Temperature-dependent $M_{SP}(T)/[dM_{SP}(T)/dT]$ (on left axis) and $\chi_0^{-1}/[(d\chi_0^{-1})/dT]$ (on right axis) plots, where black solid lines represent the straight fit using Eqs. 5 and 6, respectively. (g-i): $M$-$H$ curves at $\approx T_C$ for GdZn$_{1-x}$Ga$_x$ ($x=0-0.1$) samples, respectively. Insets show the log-log plot, where solid red lines represent the straight fit for $H \geq 2$ kOe.} 
\label{Fig6_Critical2}
\end{figure*}

We employed an iterative method to estimate the exact values of the critical exponents, $\beta$ and $\gamma$ \cite{Liu2018, Zhang2012}. Starting with the values predicted by mean field theory ($\beta = 0.5$ and $\gamma = 1$), we extracted $M_{SP}(T)$ and $\chi_0^{-1}(T)$ from the y- and x-intercepts of the Arrott plots [Figs. 1(a), 2(a), and 3(a) of \cite{SI}] and then used equations (1) and (2) to obtain a new set of $\beta$ and $\gamma$, which were subsequently used to reconstruct the MAPs. Next, extrapolating the linear fit curves in the high-field region led to another set of modified $\beta$ and $\gamma$, and consequently a new set of MAPs as well. We continued this iterative process until we obtained a set of stable values of $\beta$ and $\gamma$ within the error bars. These values of the critical exponents are independent of their initial choice, confirming the reliability of this method. The final values of $\beta$, $\gamma$, and $T_{\rm C}$ are given in Table \ref{T_critical}  for all the samples. In Figs. \ref{Fig5_Critical1}(d--f), we show the resultant MAPs constructed using these final $\beta$ and $\gamma$ values for $x = 0-0.1$ samples, respectively. All the samples exhibit linear behavior of the Arrott curves for $H >$ 10 kOe, with a slight bend in the curves in the low magnetic field region (not shown), possibly due to the complex multi-domain spin structure in this region \cite{Nehla2019}. The linear behavior of the Arrott curves indicates the correct extracted values of $\beta$ and $\gamma$, as per the Arrott-Noakes equation. 

The magnetic ordering temperature, $T_{\rm C}$, is defined as the temperature above which $M_{SP}$ vanishes and below which $\chi_0$ diverges. The linear extrapolation of the straight curves passes through the origin at $T_{\rm C}$ for all the samples [see blue line in Figs. \ref{Fig5_Critical1}(d--f) for $x = 0-0.1$, respectively], which further confirms the reliability of the extracted parameters. The best fit of $M_{SP}(T)$ and $\chi_0^{-1}(T)$ curves using equations (1) and (2) with the final values of $\beta$ and $\gamma$ are shown in Figs. \ref{Fig6_Critical2}(a--c) for the samples, respectively. The fitted curves successfully reproduce the experimental data over a wide temperature range. Note that equations (1) and (2) are defined within the asymptotic region ($\epsilon \rightarrow 0$); however, we included a slightly wider temperature range for fitting to obtain more data points and thus robust values of $\beta$ and $\gamma$. Reducing the temperature range of the fitting does not alter the final values of the critical exponents within the errors.

For further validation of the values of the critical exponents, we use the Kouvel-Fisher method, which redefines the equations of states expressed in equations (1) and (2) as: 

\begin{eqnarray}
\frac{M_{SP}(T)}{\left(\frac{dM_{SP}(T)}{dT}\right)} = \frac{T - T_C}{\beta} \\
\frac{\chi_0^{-1}(T)}{\left(\frac{d\chi_0^{-1}(T)}{dT}\right)} = \frac{T - T_C}{\gamma} 
\end{eqnarray}

In Figs. \ref{Fig6_Critical2}(d--f), we plot $M_{SP}(T)/[dM_{SP}(T)/dT]$ (on the left axis) and $\chi_0^{-1}/[d\chi_0^{-1}/dT]$ (on the right axis) vs. $T$ for the $x = 0-0.1$ samples, respectively. The values of $\beta$, $\gamma$, and $T_{\rm C}$ extracted from the slope and y-intercepts of these plots are given in Table \ref{T_critical} for all three samples, which are in good agreement with those calculated from the MAPs using the iterative method discussed above, confirming their accuracy. Furthermore, magnetization isotherms ($M$ vs. $H$) at $T_{\rm C}$ should exhibit linear behavior on a log-log scale according to equation (3). Therefore, to verify the correct values of $T_C$ extracted from the MAP and KF methods, in Figs. \ref{Fig6_Critical2}(g--i), we present the $M$ vs. $H$ curves for the $x = 0-0.1$ samples, respectively, at their respective magnetic ordering temperatures. The $M$ vs. $H$ plots at $T_{\rm C}$ in the log-log scale show linear behavior [insets of Figs. \ref{Fig6_Critical2}(g--i)], and $\delta$ for the samples was calculated from their slopes using equation (3) (Table \ref{T_critical}). Again, the critical exponent $\delta$ is related to $\beta$ and $\gamma$ through the Widom scaling relation as follows \cite{Widom1965, Kaul1985}:

\begin{table*}
    \centering
    \caption{Critical exponents extracted using different methods (this work) and their comparison with the theoretical values for the different universality classes.}
    \begin{tabular}{p{2.5cm}p{2.5cm}p{2cm}p{2cm}p{2cm}p{2cm}p{2cm}}
        \hline
        Sample/Class & Method & $\alpha$ & $\beta$ & $\gamma$ & $\delta$ & $T_C$ (K) \\
        \hline
        & MAP & -0.003 & 0.448(1) & 1.079(2) & 3.408(3) & 268.5(1) \\
        $x=0$ & KF &  & 0.443(3) & 1.079(3) & 3.436(6) & 268.6(1) \\
        & Critical isotherm &  &  &  & 3.435(5) &  \\
        \hline
        & MAP & 0.002 & 0.460(1) & 1.045(1) & 3.272(2) & 259.7(1) \\
        $x=0.05$ & KF &  & 0.456(6) & 1.051(6) & 3.305(8) & 259.8(1) \\
        & Critical isotherm &  &  &  & 3.111(8) &  \\
        \hline
        & MAP & -0.007 & 0.450(1) & 1.100(1) & 3.444(2) & 243.0(1) \\
        $x=0.1$ & KF &  & 0.447(3) & 1.100(3) & 3.461(6) & 243.1(1) \\
        & Critical isotherm &  &  &  & 3.478(5) &  \\
        \hline
        Mean field & Theory & 0 & 0.5 & 1.0 & 3.0 &  \\
        3D Heisenberg & Theory & -0.116 & 0.365 & 1.386 & 4.80 &  \\
        3D Ising & Theory & 0.009 & 0.325 & 1.24 & 4.81 &  \\
        Tricritical & Theory & 0.5 & 0.25 & 1.0 & 5.0 &  \\
        \hline
    \end{tabular}
    \label{T_critical}
\end{table*}

\begin{figure*} 
\includegraphics[width=6in]{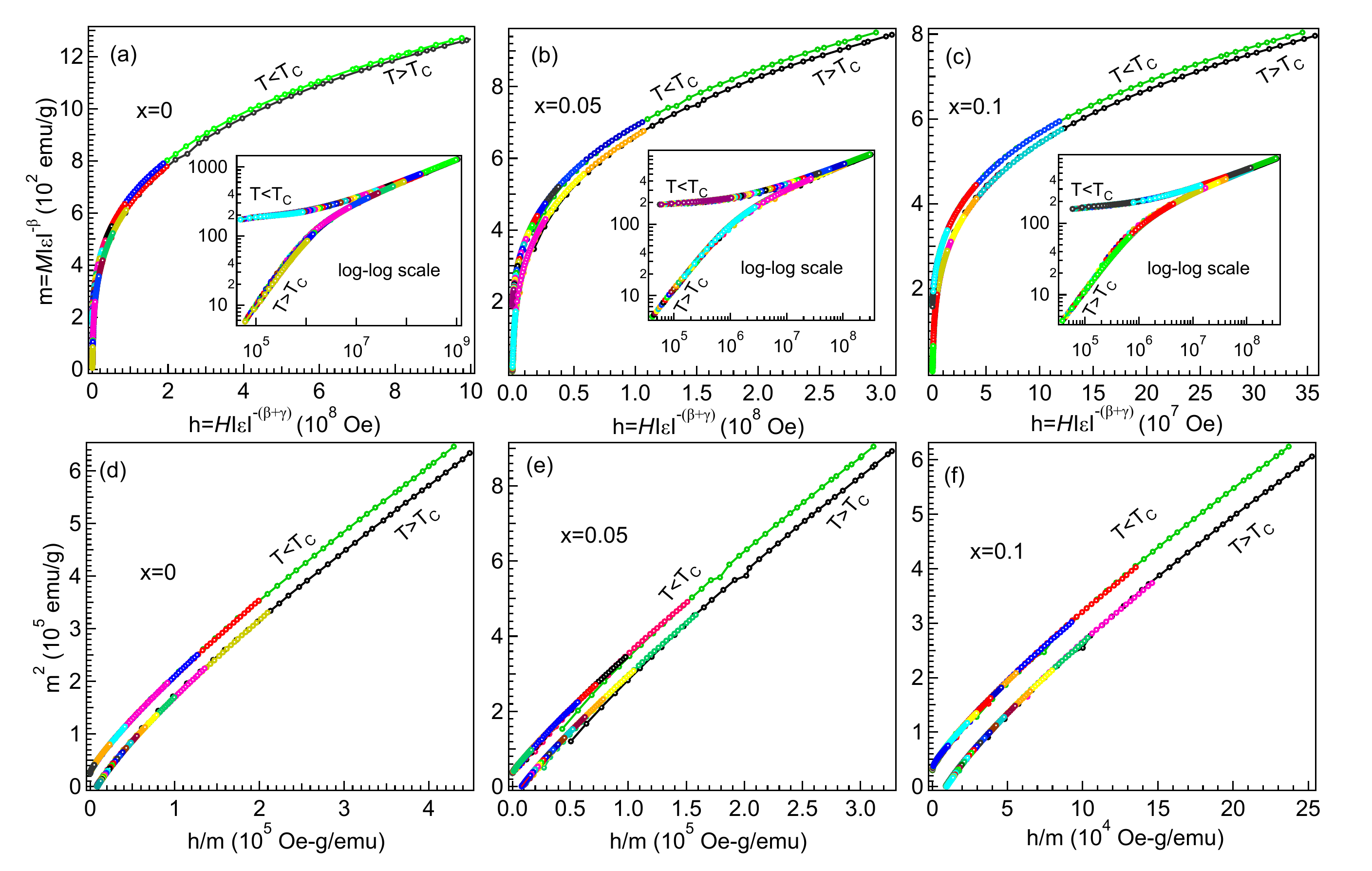}
\caption {(a-c) The reduced magnetization $m$ vs. reduced field $h$ plots for $x=0-0.1$ samples, respectively. Insets show the log-log plot of the same. (d-f) $m^2$ vs. $h/m$ plot for all the samples.} 
\label{Fig7_Critical3}
\end{figure*}

\begin{eqnarray}
 \delta = 1 + \frac{\gamma}{\beta} 
\end{eqnarray}
The values of $\delta$ were also extracted using equation (7) with the values of $\gamma$ and $\beta$ already determined. The obtained $\delta$ values via these two methods are in good agreement (see Table \ref{T_critical}). \par

Finally, we check the reliability of the critical exponents with the help of universal scaling behavior using the following equation \cite{Pramanik2009}: 

\begin{eqnarray}
M(H, \epsilon) = |\epsilon|^\beta f_\pm \left(\frac{H}{|\epsilon|^{(\beta+\gamma)}}\right) 
\end{eqnarray}

where $f_+$ and $f_-$ represent the regular functions above and below $T_{\rm C}$. For the correct values of $\beta$ and $\gamma$, the reduced magnetization $m = M |\epsilon|^{-\beta}$ vs. reduced field $h = \frac{H}{|\epsilon|^{-(\beta+\gamma)}}$ curves should collapse into two separate master curves, one above and one below $T_{\rm C}$. Figs. \ref{Fig7_Critical3}(a--c) show the $m$ vs. $h$ plots for all the samples in which the collapse of all the curves into two universal curves for all the samples is manifested. Here, we use the values of the critical exponents extracted from the MAPs. This can be more clearly observed in the log-log plots shown in their insets. Furthermore, the $m^2$ vs. $h/m$ plots [Figs. \ref{Fig7_Critical3}(d--f)] also collapse into two branches above and below $T_{\rm C}$, confirming the scaling behavior for all the samples, and hence correct values of the critical exponents $\beta$ and $\gamma$. \par

\begin{figure*} 
\includegraphics[width=6in]{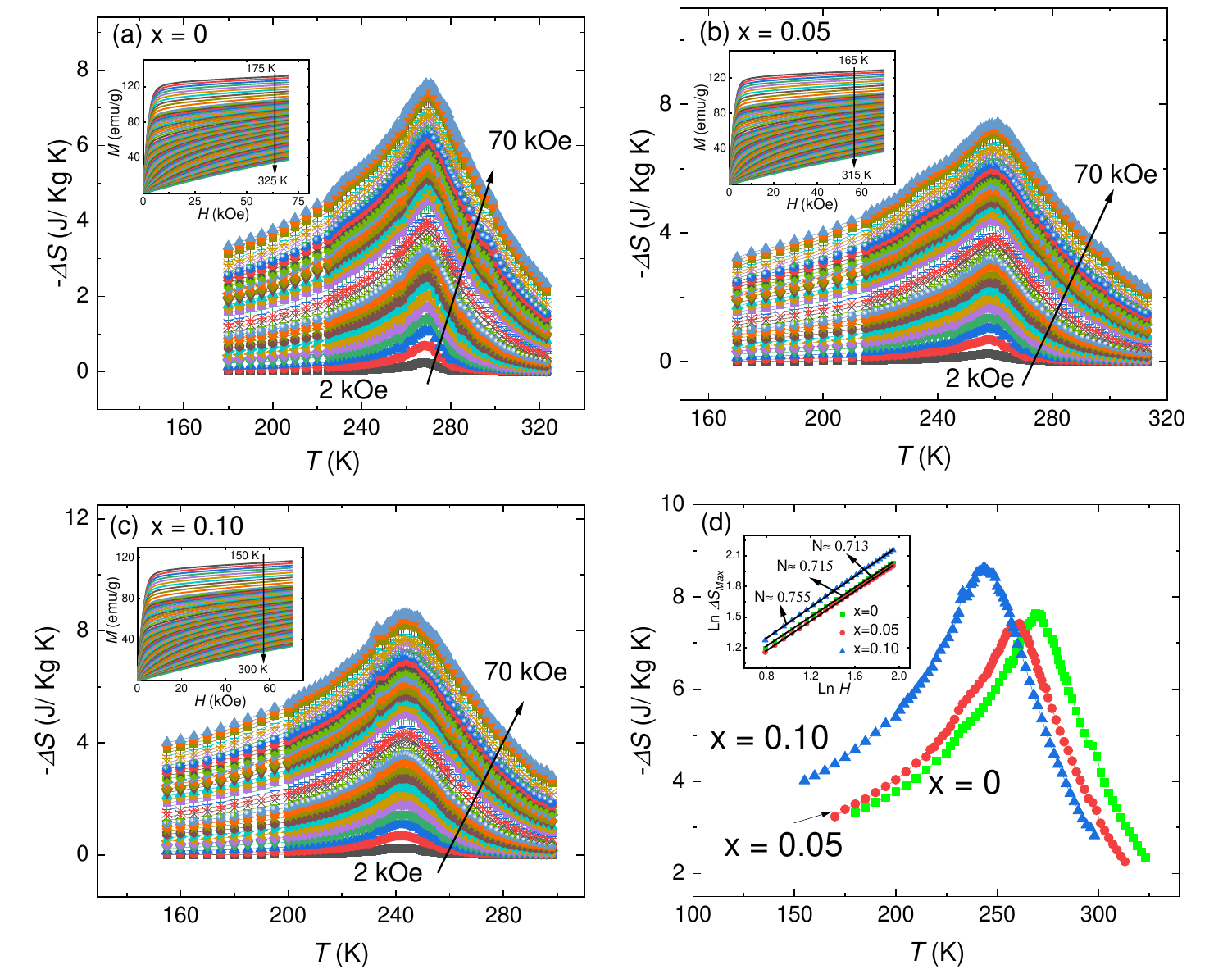}
\caption {The temperature dependence of magnetic entropy change $\Delta S$ for different magnetic field changes ranging from 2-70 kOe for GdZn$_{1-x}$Ga$_x$ with (a) $x=0$, (b) $x=0.05$, and (c) $x=0.10$. The $M$ vs. $H$ curves recorded at different temperatures for the samples are shown as insets of (a), (b), and (c). The comparison of $\Delta S(T)$ at $H=70$ kOe for the samples is shown in (d). The inset of (d) shows linear $\ln \Delta S_{\text{Max}}$ vs. $\ln H$ at $T_C$ for the samples (see equation 10). Black solid lines represent the best linear fit according to equation 10.} 
\label{Fig8_MCE}
\end{figure*}

\begin{figure} 
\includegraphics[width=3.5in]{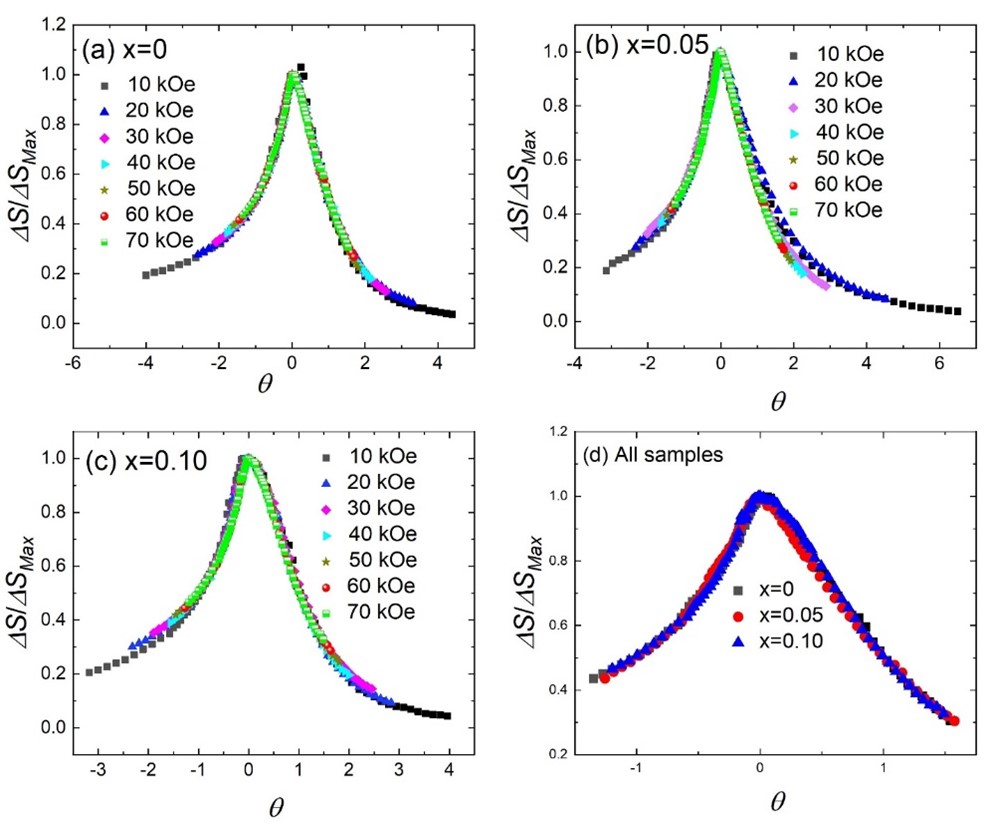}
\caption {Universal curves constructed following equation (11) for GdZn$_{1-x}$Ga$_x$: (a) $x=0$, (b) $x=0.05$, (c) $x=0.10$. (d) The universal master curve for all measured samples at $\Delta H = 70$ kOe.} 
\label{Fig9_MCE_scale}
\end{figure}

From Table \ref{T_critical}, it can be clearly inferred that for all the samples, the closest model is the mean field. Interestingly, despite a monotonic reduction in the Curie temperature with Ga substitution, our thorough critical behavior analysis indicates that long-ranged magnetic interactions and the nature of magnetic interactions remain unaltered for all the samples. However, a slight deviation towards the 3D Heisenberg model with short-range magnetic interactions makes it vital to investigate the range of magnetic interactions in these samples. As per renormalization group theory, the extended type of magnetic interactions decays as  \( J(r) \sim r^{-(d+\sigma)} \), where \( r \), \( d \), and \( \sigma \) represent spatial distance, lattice dimensionality, and range of magnetic interactions, respectively. When \( \sigma < 1.5 \), magnetic interaction is considered long-ranged, and for short-ranged interactions, \( \sigma \) becomes greater than 2. Therefore, we calculate the value of \( \sigma \) from the critical exponent \( \gamma \), extracted from the MAPs, using the following relation \cite{Fisher1972, Fisher1974, Fischer2002}: 

\begin{eqnarray}
 \gamma = 1 + \frac{4}{d} \left( \frac{n+2}{n+8} \right) \Delta \sigma + \frac{8(n+2)(n-4)}{d^2 (n+8)^2}\\
\times \left\{ 1 + \frac{2G(d/2)(7n+20)}{(n-4)(n+8)} \right\} \Delta \sigma^2 
\end{eqnarray}

where \( \Delta \sigma = (\sigma - d/2) \), \( G(d/2) = 3 - \frac{1}{4} \left( \frac{d}{2} \right)^2 \), and \( n \) represents the spin dimensionality. Considering a three-dimensional lattice and spin interactions (\( n = d = 3 \)) and using the values of \( \gamma \) from Table \ref{T_critical}, we obtain \( \sigma = 1.616 \), 1.569, and 1.644 for the \( x = 0 \), 0.05, and 0.1 samples, respectively. These values are not less than, but close to 1.5, indicating extended long-range magnetic interactions in these samples. The extracted values of \( \sigma \) can then be used to calculate the critical exponent \( \alpha \) using the relation \( \alpha = 2 - \nu d \), where \( \nu \) is the exponent of the correlation length, given as \( \nu = \gamma / \sigma \). The calculated values of the critical exponent \( \alpha \) are given in Table \ref{T_critical} and clearly indicate the predominance of mean field theory (\( \alpha \rightarrow 0 \)) in all the samples, consistent with the behavior of \( \beta \) and \( \gamma \) discussed above. Furthermore, we calculate the critical exponent \( \beta \) from the extracted values of \( \alpha \) using the relation \( \beta = (2 - \alpha - \gamma) / 2 \), which gives \( \beta = 0.462(2), 0.476(3), 0.454(2) \) for the \( x = 0, 0.05, \) and \( 0.1 \) samples, respectively. These values are close to the extracted values of \( \beta \) from the MAPs and KF methods described above, indicating the reliability of the calculated values of \( \sigma \) and \( \alpha \). The magnetic interaction in these samples decays approximately as \( J(r) \sim r^{-4.6} \), which is close to the mean field model, indicating long-range magnetic ordering in all the samples. Similar behavior has been observed in several other compounds such as CrI$_3$\cite{Liu2018}, AlCMn$_3$ \cite{Zhang2012}, PrCrGe$_3$ \cite{Yang2021}, etc. The almost same value of \( \sigma \) for all three samples also confirms that the presence of the secondary Gd(Ga,Zn)$_2$ phase in Ga-doped samples has an insignificant impact in the critical region of the phase transition. Otherwise, \( \sigma \) would differ for the samples with \( x = 0.05 \) and 0.1 compared to GdZn where this secondary phase was not observed. 

\subsection{Magnetocaloric behavior}

The virgin magnetization isotherms have been  utilized to calculate the  the temperature dependence of the magnetic entropy change, $-\Delta S$, for a given change in the magnetic field ($\Delta$\textit{H}) using  the following Maxwell’s thermodynamic relation: \cite{Phan_JMMM_07, Kumar_PRB_24}

\begin{eqnarray} 
 \Delta S (T, \Delta H)=\int_0^H\bigg (\frac{\partial M(T,H)} {\partial T}\bigg)_H dH
\end{eqnarray}

The $-\Delta$\textit{S}(\textit{T}) curves at different $\Delta$H are shown in Figs. \ref{Fig8_MCE}(a--c) for the $x =$0--0.10 samples, respectively.  The maximum values of $-\Delta S$, referred to as $\Delta S_{\text{Max}}$ hereafter, obtained in the vicinity of the transition are moderate (7-8~J/Kg K) for all the samples.  We compare the -$\Delta$\textit{S}(\textit{T}) curves at different $\Delta $\textit{H} =70 kOe for all the samples in Fig. \ref{Fig8_MCE}(d), which indicate that the Ga doping has a minimal effect on $\Delta S_{\text{Max}}$. Note that both GdZn$_2$ and GdGa$_2$ show magnetic transitions at much lower temperatures \cite{Debray1970, Tsai1979}. So one can assume that the Gd(Ga,Zn)$_2$-type impurity phase present in the Ga-doped samples would be in a paramagnetic state near the $T_{\rm C}$ of our present samples. Since we did not consider the mass of that fraction in our calculation of $\Delta S_{\text{Max}}$ per kg, the values of $\Delta S_{\text{Max}}$ are underestimated. Although the study of MCE of materials heavily emphasizes the development of energy-efficient and environmentally friendly magnetic refrigeration, the analysis of how different magnetocaloric parameters evolve with temperature and magnetic field is very useful to underpin deep insights into the complex magnetic behavior of materials, including the nature of their magnetic interactions and phase transitions \cite{Zhu2003, Haldar2010, Chandra2012, Law2018}. For the present samples, the magnetic field dependence of $\Delta S$ follows a power law, such as \cite{Franco2006, Biswas2013}: 

\begin{eqnarray}
 \Delta S \propto H^N 
\end{eqnarray}

with $N$, extracted from the slope of the linear $\ln \Delta S$ vs. $\ln H$, close to $2/3$ at $T_C$ [see inset of \ref{Fig8_MCE}(d)], in accordance with mean-field theory (MFT). The exponent $N$ is also a function of $\gamma$, $\beta$, and $\delta$ \cite{Biswas2020Jalcom}. The calculated $N$ using the critical exponents presented in Table \ref{T_critical} for the samples is close to that obtained from the slope of $\ln \Delta S$ vs. $\ln H$.

\begin{figure} 
\includegraphics[width=3.5in]{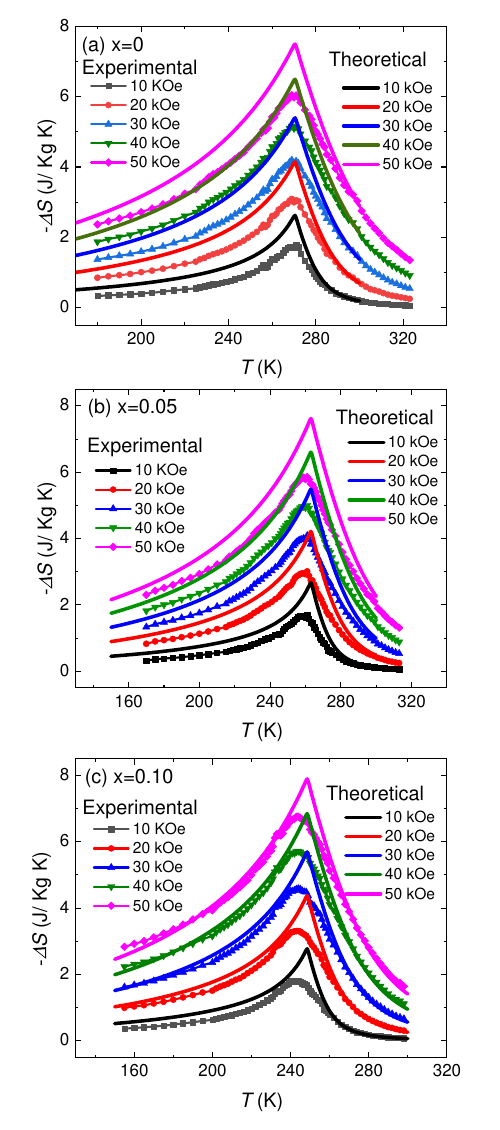}
\caption {The comparison between experimental and theoretically calculated temperature dependence of $\Delta S$ at 10, 20, 30, 40, and 50 kOe for GdZn$_{1-x}$Ga$_x$ with (a) $x=0$, (b) $x=0.05$, and (c) $x=0.1$ for the GdZn$_{1-x}$Ga$_x$ alloys under several magnetic field variations. The symbols represent experimental data, and the lines depict the mean-field calculations.} 
\label{Fig10_MCE_Theory}
\end{figure}

The magnetic field and temperature dependencies of -$\Delta S$ can also be described via the construction of a universal curve \cite{Franco2006}. It is well established that for a second-order phase transition, the $-\Delta S(T)$ dependence normalized to their peak values obtained for different magnetic fields collapse into a single master (universal) curve when the temperature axis is rescaled by introducing a new variable, $\theta$, as follows \cite{Franco2006}: 
\begin{eqnarray}
 \theta = \begin{cases} -\frac{T - T_C}{T_{\text{ref1}} - T_C} & \text{for } T \leq T_C \\ \frac{T - T_C}{T_{\text{ref2}} - T_C} & \text{for } T > T_C \end{cases} 
\end{eqnarray}
 Here, $T_{\text{ref1}}$ and $T_{\text{ref2}}$ are two reference temperatures - one below and one above $T_C$ - that satisfy the relation $\Delta S(T_{\text{ref1}})/\Delta S_{\text{Max}} = f = \Delta S(T_{\text{ref2}})/\Delta S_{\text{Max}}$. Taking $f = 0.5$, universal curves have been constructed individually for all three samples [Figs. \ref{Fig9_MCE_scale}(a--c)]. Moreover, the universal curves for all three samples can also be collapsed into a master universal curve [Fig. \ref{Fig9_MCE_scale}(d)], indicating that all the samples belong to the same universality class and they exhibit similar types of long-range magnetic interactions following mean-field theory (MFT).

\begin{figure} 
\includegraphics[width=3.5in]{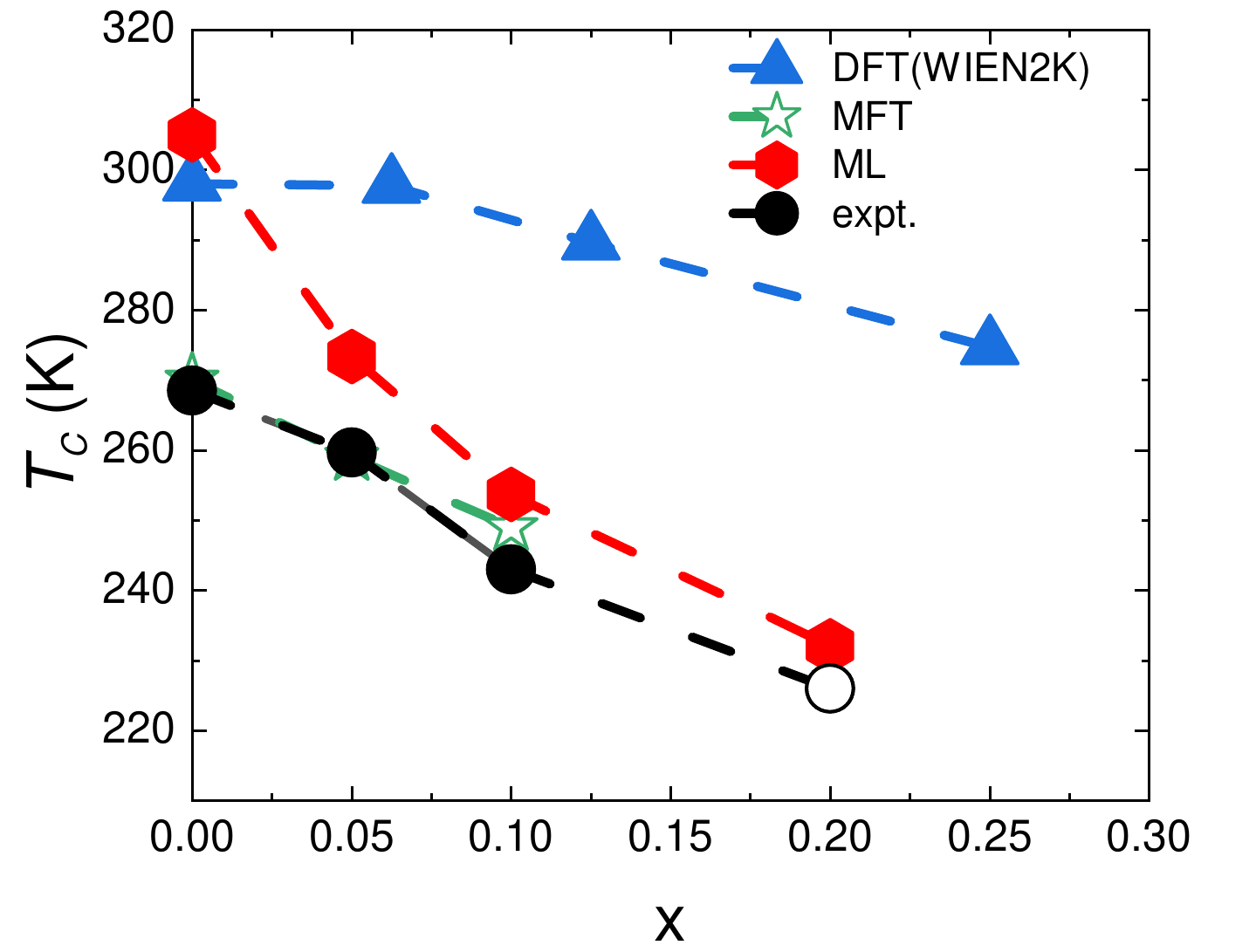}
\caption {Comparison between experimentally obtained $T_C$'s of GdZn$_{1-x}$Ga$_x$ with different $x$ and $T_C$'s of respective samples estimated by different theoretical methods: mean-field theory (MFT), machine learning based model (ML), and density functional theory calculation (DFT). The experimental $T_C$ for the sample with $x=0.2$ has been taken from ref. \cite{Petit2020}.} 
\label{Fig11_Tc_Theory}
\end{figure}

\subsection {MFT model to describe MCE}

Since the MFT is the best model for describing magnetic interactions in the vicinity of the transition for the GdZn$_{1-x}$Ga$_x$ ($x = 0 - 0.1$) samples, we formulated a theoretical model based on MFT to describe the temperature dependence of magnetic entropy change. In this context, we start from the Heisenberg Hamiltonian, with nearest-neighbors exchange interactions, in the presence of a magnetic field \cite{Gomes2006}: 

\begin{eqnarray}
 H = -j_0 \sum_{\langle ij \rangle} \mathbf{S}_i \cdot \mathbf{S}_j - g\mu_B \mu_0 H \sum_i S_i^z 
\end{eqnarray}

Within the Weiss molecular-field approach, this simplifies to the action of an effective magnetic field on the Gd spins: 

\begin{eqnarray}
 H = -h S^z \quad  
\end{eqnarray}

where \( h = zj_0 \langle S_z \rangle + g\mu_B \mu_0 H \).

In this approach, the transition temperature is related to the exchange interaction by the equation: 

\begin{eqnarray}
 T_{\rm C} = \frac{zj_0 S(S+1)}{3k_B} 
\label{eq_j0}
\end{eqnarray}

In this relation, \( z \) refers to the coordination number (number of magnetic nearest neighbors) of a given Gd-site. For the studied compounds, the introduction of Ga in the matrix does not affect this number, because the crystal structure does not change for the considered range of \( x \) \cite{Blaha2001, Subhan2020}. Therefore, the observed change in   $T_{\rm C}$ is related to the variation of the coupling between the localized spins of Gd and the conduction electrons \cite{Gignoux1991}. Following Guo et al. \cite{Guo2000},   $T_{\rm C}$(x) can be expressed as: 

\begin{eqnarray}
 T_ {\rm C}(x) = T_C^0 (1 + ax) 
\label{eq_Tc}
\end{eqnarray}

where \( a \) is a material-dependent parameter. Using our experimentally obtained \( T_{\rm C} \)'s, we noted \( a = 1/T_{\rm C}^0 \), \( (dT_{\rm C})/dx = -0.790 \), and an RMSE of 1.63 K. We compare the $T_{\rm C}$'s for all three samples, both experimental and those expected from relation \ref{eq_Tc} in Fig. \ref{Fig11_Tc_Theory}. We also list the exchange couplings obtained from equation \ref{eq_j0} in Table \ref{T_theory}. The isothermal entropy was calculated using the relation: 

\begin{eqnarray}
 \Delta S_T = S_{\text{mag}}(T, H=0) - S_{\text{mag}}(T, H \neq 0)  
\end{eqnarray}

The magnetic entropy (\( S_{\text{mag}} \)) as a function of temperature and magnetic field is given by: 

\begin{eqnarray}
 S_{\text{mag}}(T, H) = k_B T \left( \ln \sum_{\epsilon_i} e^{-\beta \epsilon_i} + \frac{\beta \left( \sum_{\epsilon_i} \epsilon_i e^{-\beta \epsilon_i} \right)} { \sum_{\epsilon_i} e^{-\beta \epsilon_i} }\right) 
\label{eq_delS}
\end{eqnarray}

where \( k_B \) is the Boltzmann constant, \( \beta = 1/k_B T \), and \( \epsilon_i \) are the eigenvalues from Hamiltonian (1) which carry the dependence on the effective field \( h \). We calculated \( -\Delta S(T) \) using equation \ref{eq_delS} and compared the results with the experimentally obtained \( -\Delta S(T) \) as shown in Fig. \ref{Fig10_MCE_Theory} (a--c) for the $x =$ 0, 0.05, and 0.1 samples, respectively. There exists reasonably good agreement between the experimental data and the results of the mean-field calculations for the samples, although the experimental values are expectedly lower due to the presence of weakly magnetic impurities and the polycrystalline nature of the samples as observed in earlier studies \cite{Alho2022, Biswas2020Jalcom}. Since the theoretical model based on MFT can describe the experimentally obtained magnetocaloric behavior, one can safely conclude that there exists long-range magnetic ordering in GdZn$_{1-x}$Ga$_x$, but the magnetic exchange weakens with an increase in Ga concentration.

\subsection{DFT and ML based theory}

\begin{table}
    \centering
    \caption{The calculated Gd-Gd interaction parameters  according to Eq. 14 for GdZn$_{1-x}$Ga$_x$.}
    \begin{tabular}{p{0.8cm}p{1.5cm}p{1.8cm}p{1.8cm}p{1.8cm}}
        \hline
        $x$ & $J_0$ (meV): MFT & $J_{12}^{RKKY}$ (r$_{Gd1-Gd2}$) & $J_{13}^{RKKY}$ (r$_{Gd1-Gd3}$) & $J_{14}^{RKKY}$ (r$_{Gd1-Gd4}$) \\
        \hline
        0.0 & 0.74 & -0.0019762 & 0.0099488 & -0.001649 \\
        0.05 & 0.72 & 0.000196 & 0.0108007 & -0.002686 \\
        0.1 & 0.68 & 0.0020608 & 0.0098478 & -0.001994 \\
        \hline
    \end{tabular}
    \label{T_theory}
\end{table}

In an earlier study, the  $T_{\rm C}$'s for GdZn$_{1-x}$Ga$_x$ were calculated using DFT based on disordered local moment (DLM) theory combined with the self-interaction corrected local spin density approximation (SIC-LSDA) \cite{Petit2020}. The compositional dependence of  $T_{\rm C}$ showed an approximately linear decreasing trend for $x > 0.1$, but there was a prediction of deviation from that trend for lower concentrations of Ga. However, our present experimental results conclusively establish that  $T_{\rm C}$ for the samples with $x \leq 0.1$ also decreases. Here we address this discrepancy using different theoretical techniques. First, we calculate  $T_{\rm C}$ by using a FP-LAPW)method \cite{Blaha2001}. The Curie temperature of the given compositions is calculated using energy analysis within various magnetic configurations, as explained in the earlier study using the classical spin Heisenberg model \cite{Subhan2020}. The exchange energy parameter (\( J \)) can be obtained by the relation \( J = \frac{\Delta E}{NM^2} \), where \( N \) and \( M \) are the total number of magnetic atoms per unit cell and the magnetic moment of a single magnetic atom in the unit cell, respectively, and \( \Delta E \) is the difference between AFM and FM configuration energies, which relates directly to the Curie temperature (i.e., \( T_{\rm C} = \frac{2}{3} \frac{\Delta E_{\text{AFM-FM}}}{N k_B} \)) \cite{Subhan2020}. As shown in Fig. \ref{Fig11_Tc_Theory}, the compositional dependence of $T_{\rm C}$  shows a gradual and steady decrease across the whole studied compositional range despite the estimated values of  $T_{\rm C}$'s are considerably larger than those obtained via experiment.\par

\begin{figure} 
\includegraphics[width=3.5in]{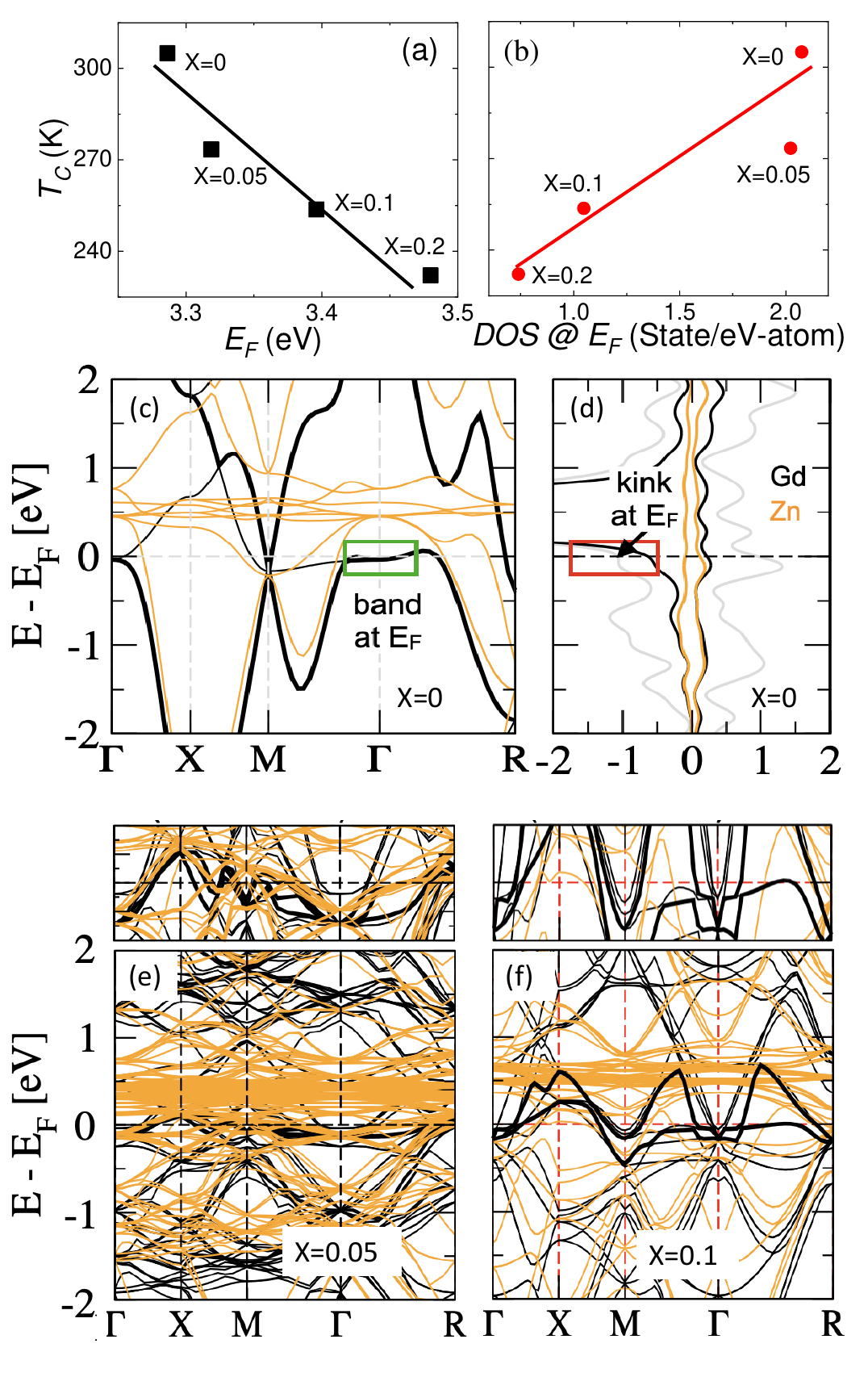}
\caption {The comparison between electronic structure quantities (calculated using VASP based DFT) such as (a) electronic filling (Fermi-level, $E_F$ in eV), and (b) density of states at $E_F$ (states/eV-atom) showing systemetic change with T$_{C}$. The dotted connecting lines are as guide for eyes. The electronic band structure of GdZn$_{1-x}$Ga$_x$ in the Pm-3m phase at (c)and (d) $x = 0.0$, (e) $x = 0.05$, and (f) $x = 0.10$. The magnified scale (top panel) of the band structure in (d) and (e) highlights the reduced complexity in the band structure at or near the Fermi level with Ga concentration.} 
\label{Fig12_DOS}
\end{figure}

Recently, a physics-informed machine-learning (ML) model has been developed to predict  $T_{\rm C}$ of rare earth materials with reasonably low absolute error \cite{Singh2023}. We calculated  $T_{\rm C}$'s for the present samples using that model with no additional training. The estimated  $T_{\rm C}$'s for Ga-doped samples are in good agreement with experiments (see Fig. \ref{Fig11_Tc_Theory}), although there is a noticeable overestimation of $T_{\rm C}$ for GdZn. It appears that the agreement between ML estimation and experimental values improves with the increase of Ga concentration (see Fig. \ref{Fig11_Tc_Theory}). In earlier studies, Singh et al. \cite{Singh2023} and Deng et al. \cite{Deng2018} explored the correlation between  $T_{\rm C}$ and density of electronic states at the Fermi level for rare-earth intermetallics. According to their studies,  $T_{\rm C}$ can be tuned by the chemical alloying route to modulate both electron filling and densities of states (DOS) at the Fermi level. We performed a DFT study on a special quasi-random supercell (SQS) generated from disordered supercells using a Monte Carlo scheme to randomly distribute atoms and calculate electronic filling at the Fermi level, $E_{\text{F}}$, and density of states at the Fermi level, DOS@E$_{F}$. In Figs. \ref{Fig12_DOS}(a) and (b)  we found a systematic change in electronic filling of states near the Fermi level and DOS@E$_{F}$ upon alloying with Ga in GdZn$_{1-x}$Ga$_x$ when plotted with respect to the ML-estimated Curie temperature. This systematic change was attributed to Ga alloying, which effectively introduces more electrons into the system, leading to the filling of previously unoccupied electronic states at the Fermi level. We believe that this filling of electronic states at E$_{F}$ shifts the hybridized bands to lower energy values below the Fermi level, as reflected in the increasing trend in the magnitude of Fermi energy [Fig. \ref{Fig12_DOS}(a)] or decreasing trends in DOS@E$_{F}$ [Fig. \ref{Fig12_DOS}(b)] in GdZn$_{1-x}$Ga$_x$.\par

Earlier, Zhao et al. \cite{Zhao2021} performed first-principles calculations on the pure GdZn compound, revealing its complex electronic band structures. To investigate the effect of Ga-doping on the electronic band structure, we evaluated the band structures of GdZn$_{1-x}$Ga$_x$ (x=0, 0.05, and 0.1) using density functional theory (DFT) methods (VASP)\cite{Kresse1994, Hafner2008}. The electronic structure of GdZn is shown in Fig. \ref{Fig12_DOS}(c), where the band structure (left panel) exhibits flat-band features near the Fermi level along the $\Gamma$-R and $\Gamma$-M high-symmetry points. Such features in electronic bands are known to induce increased ferromagnetic behavior and high Curie temperature due to changes in Hubbard repulsion \cite{Bobrow2021, Maksymenko2012}. Notably, we also observed a small kink in the total density of states (DOS) in the down-spin channel, which is sometimes correlated with possible competing instabilities \cite{Singh2023}. However, these instabilities in the form of flat bands and/or kinks in DOS completely disappear upon the addition of Ga, as shown by the magnified band structures in the top panels of Figs. \ref{Fig12_DOS}(d, e). We believe that the complexity associated with the flat bands in GdZn makes it challenging to accurately position these energy bands, which directly impacts the accuracy of the total energy theoretically calculated for this compound. Even in the ML model, different key physics-based alloy-specific features such as $f$-bandwidth, bond length, and structural properties, directly calculated from DFT, were used, connecting it with the band structure. As a result, the discrepancy between the ML estimated $T_{\rm C}$ and experiment is highest in the case of GdZn, which has the most complex band structure (with competing instabilities in the electronic structure) among our alloy series. These instabilities are absent in the electronic structure of Ga-doped samples, which do not exhibit flat bands. Consequently, the ML-calculated  $T_{\rm C}$'s for those samples are in good agreement with the experiments.\par

Finally, to gain more insights into the magnetic indirect exchange interactions between Gd atoms, we estimate the exchange interaction parameters between the i$^{\rm th}$ and j$^{\rm th}$ atoms using a model in accordance with the RKKY theory \cite{Priour2004}: 

\begin{eqnarray}
J_{ij}^{RKKY} = J_0 r_{ij}^4 \left[ \sin(2k_F r_{ij}) - 2k_F r_{ij} \cos(2k_F r_{ij}) \right]
\end{eqnarray}

where \( r_{ij} = \vec{R}_i - \vec{R}_j \) is the spatial separation between the magnetically coupled Gd atoms; \( k_F \sim \left( \frac{3\pi^2 nZ}{V_0^3} \right)^{1/3} \), where \( V_0 \) is the cell volume, \( n \) is the number of atoms in the unit cell, and \( Z \) is the average number of valence electrons per atom. Considering the values of \( J_0 \) calculated from MFT (first column of Table \ref{T_theory}) between neighboring Gd atoms (\( J_{12}^{RKKY}, J_{13}^{RKKY}, \) and \( J_{14}^{RKKY} \) etc.) are given in Table \ref{T_theory}. Interestingly, while \( J_{12}^{RKKY}, J_{13}^{RKKY}, \) and \( J_{14}^{RKKY} \) follow the usual sinusoidal variation as per the classical RKKY model for the ordered GdZn, such behavior is absent for Ga-doped samples. This clearly points out that the change in the number of electrons due to the introduction of  Ga at the Zn-site impacts the indirect exchange interaction, which is reflected in the significant change in T$_{C}$  for the GdZn$_{1-x}$Ga$_x$ with Ga -doping.

\section{\noindent ~Conclusion}

The introduction of Ga at the Zn site of GdZn has a considerable impact on the magnetic properties and electronic structure. With an increase in Ga concentration, the magnetic transition temperature decreases but long-ranged ferromagnetic ordering stabilizes. The decrease of transition temperature with Ga doping qualitatively agrees with  a ML-based model. The analysis of critical exponents indicates that mean-field theory (MFT) is applicable to describe the magnetic behavior of GdZn$_{1-x}$Ga$_x$ in the vicinity of the transition.  The samples exhibit moderate magnetocaloric effect as expected for materials with broad second-order transition. A theoretical model is developed based on MFT, which successfully describes the temperature dependence of the magnetocaloric response of the alloy series. On the other hand, our DFT calculations reveal complex electronic band structure of GdZn, namely a flat band, a signature of instability near Fermi level.  Interestingly, this flat band signature is absent in the Ga-doped samples. Moreover, an ML-based model accurately describes the experimental trend of  $T_{\rm C}$ changing with Ga concentration, and the agreement between theoretical and experimental values of  $T_{\rm C}$ improves with an increase in Ga concentration.

\subsection{Acknowledgments}

 This work was performed at Ames National Laboratory and supported by the Division of Materials Science and Engineering of the Office of Basic Energy Sciences, Office of Science of the U.S. Department of Energy (DOE). Ames Laboratory is operated for the U.S DOE by Iowa State University under Contract No. DE-AC02-07CH11358. BCM, POR, BPA, VSRS, EPN and PJVR acknowledge the financial support of Coordenação de Aperfeiçoamento de Pessoal de Nível Superior – Brasil (CAPES) – Finance Code 001, CNPq – Conselho Nacional de Desenvolvimento Científico e Tecnológico – Brazil and FAPERJ - Fundação de Amparo à Pesquisa do Estado do Rio de Janeiro. Authors would like to thank D. Jing of Materials Analysis and Research Laboratory, Iowa State University, for acquisition of XPS and EDS data.


\begin{thebibliography}{99}

\bibitem{Szytula1993}A.~Szytuła, Magnetism of rare earth intermetallics, Phys. Scr. \textbf{1993}, 284 (1993).

\bibitem{Gignoux1991}D.~Gignoux and D.~Schmitt, Rare earth Intermetallics, J. Magn. Magn. Mater. \textbf{100}, 99 (1991).

\bibitem{Savchenkov2023}P.~S.~Savchenkov and P.~A.~Alekseev, Uncommon Magnetism in Rare-Earth Intermetallic Compounds with Strong Electronic Correlations, Crystals \textbf{13}, 1238 (2023).

\bibitem{Chinchure2002}A.~D.~Chinchure, E.~M.~Sandoval, and J.~A.~Mydosh, Metamagnetism and giant magnetoresistance of the rare-earth intermetallic compounds R$_2$Ni$_2$Pb (R=Er, Ho, Dy), Phys. Rev. B \textbf{66}, 020409(R) (2002).

\bibitem{Chakraborty2024}W. Simeth, M. C. Rahn, A. bauer, M. Meven, and C. Pfleiderer, Topological aspects of multi-k antiferromagnetism in cubic rare-earth compounds, J. Phys.:cond.Mat. \textbf{36}, 215602 (2024). 

\bibitem{Biswas2024a}A.~Biswas, R.~K.~Chouhan, O.~Dolotko, P.~Manfrinetti, S.~Lapidus, D.~L.~Schlagel, and Y.~Mudryk, Exceptional magnetic and magnetoelastic behavior of rare-earth non-centrosymmetric Sm$_7$Pd$_3$, Acta Mater. \textbf{265}, 119630 (2024).

\bibitem{Yu2020}J.~Bouaziz, G.~Bihlmayer, C.~Patrick, J.~Staunton, and S.~Blugel, Origin of incommensurate magnetic order in the RALSi magnetic Weyl Semimetals (R =Pr, Nd, Sm) Phys. Rev. B \textbf{109}, L201108 (2024).

\bibitem{Biswas2020PRB}A.~Biswas, N.~A.~Zarkevich, A.~K.~Pathak, O.~Dolotko, I.~Z.~Hlova, A.~V.~Smirnov, Y.~Mudryk, D.~D.~Johnson, and V.~K.~Pecharsky, First-order magnetic phase transition in Pr$_2$In with negligible thermomagnetic hysteresis, Phys. Rev. B \textbf{101}, 224402 (2020).

\bibitem{Alho2022}B.~P.~Alho, P.~O.~Ribeiro, V.~S.~R.~de Sousa, B.~C.~Margato, A.~Biswas, T.~Del Rose, R.~S.~de Oliveira, E.~P.~Nóbrega, P.~J.~von Ranke, Y.~Mudryk, and V.~K.~Pecharsky, Fathoming the anisotropic magnetoelasticity and magnetocaloric effect in GdNi, Phys. Rev. B \textbf{106}, 184403 (2022).

\bibitem{Ribeiro2022}P.~O.~Ribeiro, B.~P.~Alho, R.~S.~De Oliveira, E.~P.~Nóbrega, V.~S.~R.~de Sousa, P.~J.~von Ranke, A.~Biswas, M.~Khan, Y.~Mudryk, and V.~K.~Pecharsky, Magnetothermal properties of Ho$_{1-x}$Dy$_x$Al$_2$ (x=0, 0.05, 0.10, 0.15, 0.25, and 0.50) compounds, J. Magn. Magn. Mater. \textbf{544}, 168705 (2022).

\bibitem{Biswas2020Jalcom}A.~Biswas, T.~Del Rose, Y.~Mudryk, P.~O.~Ribeiro, B.~P.~Alho, V.~S.~R.~de Sousa, E.~P.~Nóbrega, P.~J.~von Ranke, and V.~K.~Pecharsky, Hidden first-order phase transitions and large magnetocaloric effects in GdNi$_{1-x}$Co$_x$, J. Alloys Compd. \textbf{897}, 163116 (2020).

\bibitem{Biswas2022} T. Kurumaji, S. Fang, L. Ye, S. Kitou, and J. G. Checkelsky, Metamagnetic multiband Hall effect in Ising antiferromagnet ErGa$_{2}$, Proc. Nat. Acc. Sci.. (PNAS) \textbf{121}, e2318411121 (2024).

\bibitem{Ribeiro2024}W. Liu, T. Gottschall, F. Scheibel, E. Bykov, A. Aubert, N. Fortunato, B. Beckmann, A. M. Doring, H. Zhang, K. Skokov, O. Gutfleisch, A matter of performence and criticality: A review of rare-earth-based magnetocaloric intermetallic compounds, for hydrogen liquification,  J. Alloys Compd. \textbf{995}, 174612 (2024).
\bibitem{DelRose2024} T.~Del Rose, R.~Choudhary, Y.~Mudryk, D.~Haskel, A.~K.~Pathak, G.~Bhaskar, J.~V.~Zaikina, D.~D.~Johnson, and V.~K.~Pecharsky, Interplay between Kondo and magnetic interactions in Pr$_{0.75}$Gd$_{0.25}$ScGeH, J. Alloys Compd. \textbf{966}, 171351 (2024).

\bibitem{Petit2015} L.~Petit, D.~Paudyal, Y.~Mudryk, K.~A.~Gschneidner, Jr., V.~K.~Pecharsky, M.~Lüders, Z.~Szotek, R.~Banerjee, and J.~B.~Staunton, Complex Magnetism of Lanthanide Intermetallics and the Role of their Valence Electrons: Ab Initio Theory and Experiment, Phys. Rev. Lett. \textbf{115}, 207201 (2015).

\bibitem{Campbell1972}I.~A.~Campbell, Indirect exchange for rare earths in metals, J. Phys. F: Metal Phys. \textbf{2}, L47 (1972).

\bibitem{Haskel2007}D.~Haskel, Y.~B.~Lee, B.~N.~Harmon, Z.~Islam, J.~C.~Lang, G.~Srajer, Y.~Mudryk, K.~A.~Gschneidner, and V.~K.~Pecharsky, Role of Ge in Bridging Ferromagnetism in the Giant Magnetocaloric $Gd_5(Ge_{1-x}Si_x)_4$ Alloys, Phys. Rev. Lett. \textbf{98}, 247205 (2007).

\bibitem{Guillou2018}F.~Guillou, A.~K.~Pathak, D.~Paudyal, Y.~Mudryk, F.~Wilhelm, A.~Rogalev, and V.~K.~Pecharsky, Non-hysteretic first-order phase transition with large latent heat and giant low-field magnetocaloric effect, Nat. Commun. \textbf{9}, 2925 (2018).

\bibitem{Guillou2020}F.~Guillou, H.~Yibole, R.~Hamane, V.~Hardy, Y.~B.~Sun, J.~J.~Zhao, Y.~Mudryk, and V.~K.~Pecharsky, Crystal structure and physical properties of Yb$_2$In and Eu$_{2-x}$Yb$_x$In alloys, Phys. Rev. Mater. \textbf{4}, 104402 (2020).

\bibitem{Pecharsky1997}V.~K.~Pecharsky and K.~A.~Gschneidner, Jr., Giant magnetocaloric effect in Gd$_5$(Si$_2$Ge$_2$), Phys. Rev. Lett. \textbf{78}, 4494 (1997).

\bibitem{Levin2001}E.~M.~Levin, V.~K.~Pecharsky, and K.~A.~Gschneidner, Jr., Spontaneous generation of voltage in Gd$_5$(Si$_x$Ge$_{4-x}$) during a first-order phase transition induced by temperature or magnetic field, Phys. Rev. B \textbf{63}, 174110 (2001).

\bibitem{Nazih2003}M.~Nazih, A.~de Visser, L.~Zhang, O.~Tegus, and E.~Brück, Thermal expansion of the magnetorefrigerant Gd$_5$(Si,Ge)$_4$, Solid State Commun. \textbf{126}, 255 (2003).

\bibitem{Chakraborty2022}S.~Chakraborty, S.~Gupta, S.~Pakhira, R.~Choudhary, A.~Biswas, Y.~Mudryk, V.~K.~Pecharsky, D.~D.~Johnson, and C.~Mazumdar, Ground-state degeneracy and complex magnetism of geometrically frustrated Gd$_2$Ir$_{0.97}$Si$_{2.97}$, Phys. Rev. B \textbf{106}, 224427 (2022).

\bibitem{Velez2010}S.~Velez, J.~M.~Hernandez, A.~Fernandez, F.~Macià, C.~Magen, P.~A.~Algarabel, J.~Tejada, and E.~M.~Chudnovsky, Magnetic deflagration in Gd$_5$Ge$_4$, Phys. Rev. B \textbf{81}, 064437 (2010).


\bibitem{Morellon1998}L.~Morellon, J.~Stankiewicz, B.~García-Landa, P.~A.~Algarabel, and M.~R.~Ibarra, Giant magnetoresistance near the magnetostructural transition in Gd$_5$(Si$_{1.8}$Ge$_{2.2}$), Appl. Phys. Lett. \textbf{73}, 3462 (1998).

\bibitem{DelRose2021}T.~Del Rose, A.~K.~Pathak, Y.~Mudryk, and V.~K.~Pecharsky, Distinctive exchange bias and unusual memory effects in magnetically compensated Pr$_{0.75}$Gd$_{0.25}$ScGe, J. Mater. Chem. C \textbf{9}, 181 (2021).

\bibitem{Kurumaji2019}T.~Kurumaji, T.~Nakajima, M.~Hirschberger, A.~Kikkawa, Y.~Yamasaki, H.~Sagayama, H.~Nakao, Y.~Taguchi, T.~Arima, and Y.~Tokura, Skyrmion lattice with a giant topological Hall effect in a frustrated triangular-lattice magnet, Science \textbf{365}, 914 (2019).

\bibitem{Kanematsu1969}K.~Kanematsu, G.~T.~Alfieri, and E.~Banks, Magnetic Studies of Rare Earth Zinc Compounds with CsCl Structure, J. Phys. Soc. Jpn. \textbf{26}, 244 (1969).

\bibitem{Oppelt1972}A.~Oppelt, E.~Dormann, and K.~H.~J.~Buschow, NMR Study of Ferromagnetic Gd Intermetallic Compounds with CsCl Structure, Phys. Stat. Sol. B \textbf{51}, 275 (1972).

\bibitem{Alfieri1966}G.~T.~Alfieri, E.~Banks, and K.~Kanematsu, Magnetic Studies of Gadolinium Compounds with the CsCl Structure, J. Appl. Phys. \textbf{37}, 1254 (1966).

\bibitem{Takei1979}K.~Takei, Y.~Ishikawa, N.~Watanabe, and K.~Tajima, Magnetic Structures of GdCu$_{1-x}$Zn$_x$ System, J. Phys. Soc. Jpn. \textbf{47}, 88 (1979).

\bibitem{Yashiro1976}T.~Yashiro, Y.~Hamaguchi, and H.~Watanabe, Magnetic Structure of TbCu$_{1-x}$Zn$_x$, J. Phys. Soc. Jpn. \textbf{40}, 63 (1976).

\bibitem{Kobler1981}U.~Köbler, W.~Kinzel, and W.~Zinn, Magnetic phase diagram of GdAg$_{1-x}$Zn$_x$, J. Magn. Magn. Mater. \textbf{25}, 124 (1981).

\bibitem{Kissell1966}F.~Kissell and W.~E.~Wallace, Magnetic characteristics of some I:I compounds of the lanthanides with gold and aluminum, J. Less-Common Met. \textbf{11}, 417 (1966).

\bibitem{Sekizawa1966}K.~Sekizawa and K.~Yasukochi, Magnetic Properties of Rare Earth Intermetallic Compounds in Gd(Ag, Cd, In) and Gd(Cu, Ag, Au) Systems, J. Phys. Soc. Jpn. \textbf{21}, 684 (1966).

\bibitem{Rouchy1981}J.~Rouchy, P.~Morin, and E.~du T.~de Lacheisserie, Magnetic and magnetoelastic properties of GdZn single crystals, J. Magn. Magn. Mater. \textbf{23}, 59 (1981).

\bibitem{Petit2020}L.~Petit, Z.~Szotek, D.~Paudyal, A.~Biswas, Y.~Mudryk, V.~K.~Pecharsky, and J.~B.~Staunton, Magnetic structure of selected Gd intermetallic alloys from first principles, Phys. Rev. B \textbf{101}, 014409 (2020).

\bibitem{Sekizawa1983}K.~Sekizawa, T.~Watanabe, and K.~Yasukochi, Structural transformation and magnetic properties of (Gd$_{1-x}$R$_x$)Tl and Gd(Tl$_{1-y}$M$_y$) systems, J. Magn. Magn. Mater. \textbf{31-34}, 181 (1983).

\bibitem{Susilo2014}R.~A.~Susilo, M.~Cadogan, D.~H.~Ryan, N.~R.~Lee-Hone, R.~Cobas, and S.~Muñoz-Pérez, Spin-reorientation in GdGa, Hyperfine Interact. \textbf{226}, 257 (2014).

\bibitem{Baenziger1961}N.~C.~Baenziger and J.~L.~Moriarty Jr., Gadolinium and dysprosium intermetallic phases. II. Laves phases and other structure types, Acta Crystallogr. \textbf{14}, 948 (1961).

\bibitem{Delfino1983}S.~Delfino, A.~Saccone, and R.~Ferro, Phase Equilibria in the Gd-In and Gd-Tl Systems, Zeitschrift für Metallkunde \textbf{74}, 674 (1983).

\bibitem{Buschow1965}K.~H.~J.~Buschow, Rare earth-aluminium intermetallic compounds of the form RAl and R$_3$Al$_2$, J. Less-Common Met. \textbf{8}, 209 (1965).

\bibitem{Zhao2021}N.~Zhao, K.~Liu, and Zhong-Yi Lu, Large anomalous Hall effect induced by gapped nodal lines in GdZn and GdCd, Phys. Rev. B \textbf{103}, 205104 (2021).

\bibitem{AmesLab}Ames Laboratory, Materials Preparation Center, \url{https://www.ameslab.gov/dmse/materials-preparation-center}.

\bibitem{Holm2004}A.~P.~Holm, V.~K.~Pecharsky, K.~A.~Gschneidner Jr., R.~Rink, and M.~N.~Jirmanus, X-ray powder diffractometer for in situ structural studies in magnetic fields from 0 to 35 kOe between 2.2 and 315 K, Rev. Sci. Instrum. \textbf{75}, (2004).

\bibitem{Toby2013}B.~H.~Toby and R.~B.~von Dreele, GSAS-II: The genesis of a modern open-source all purpose crystallography software package, J. Appl. Crystallogr. \textbf{46}, 544 (2013).

\bibitem{Kouvel1964}J.~S.~Kouvel and M.~E.~Fisher, Detailed magnetic behavior of nickel near its Curie point, Phys. Rev. \textbf{136}, 1626 (1964).

\bibitem{NevesBez2018}H.~Neves Bez, H.~Yibole, A.~Pathak, Y.~Mudryk, and V.~K.~Pecharsky, Best practices in evaluation of the magnetocaloric effect from bulk magnetization measurements, J. Magn. Magn. Mater. \textbf{458}, 301 (2018)

\bibitem{Hafner2008}J.~Hafner, Ab-Initio Simulations of Materials Using VASP: Density-Functional Theory and Beyond, J. Comput. Chem. \textbf{29}, 2044 (2008).

\bibitem{Blaha2020}P.~Blaha, K.~Schwarz, F.~Tran, R.~Laskowski, G.~K.~H. Madsen, and L.~D. Marks, WIEN2k: An APW+lo program for calculating the properties of solids, J. Chem. Phys. \textbf{152}, 074101 (2020).

\bibitem{Kresse1}G.~ Kresse, and J. ~Hafner, Ab initio molecular dynamics for liquid metals, Phys. Rev. B \textbf{47}, 558(1993).

\bibitem{Monkhorst1}H.~Monkhorst, and J.~Pack, Special points for Brillouin-zone integrations, Phys. Rev. B \textbf{13}, 5188 (1976).

\bibitem{PS1} P.~ Singh, M.~Harbola, M.~Hemanadhan, A.~Mookerjee, and D.D. Johnson, Better band gaps with asymptotically corrected local exchange potentials, Phys Rev B \textbf{93}, 085204 (2016). 

\bibitem{PS2} P.~  Singh, M.K. Harbola, B. Sanyal, and A. Mookerjee, Accurate determination of band gaps within density functional formalism, Phys Rev B \textbf{87} , 235110 (2013).

\bibitem{Perdew1996}J.~P. Perdew, K.~Burke, and M.~Ernzerhof, Generalized gradient approximation made simple, Phys. Rev. Lett. \textbf{77}, 3865 (1996).

\bibitem{Dudarev1}7.	S.L. Dudarev, G.A. Botton, S.Y. Savrasov, C.J. Humphreys, and A.P. Sutton, Electron-energy-loss spectra and the structural stability of nickel oxide: An LSDA+U study, Phys. Rev. B \textbf{57}, 1505 (1998).

\bibitem{VanDeWalle2002}A.~Van De Walle, M.~Asta, and G.~Ceder, The Alloy Theoretic Automated Toolkit: A User Guide, Calphad \textbf{26}, 539 (2002).

\bibitem{Singh2023}P.~Singh, T.~Del Rose, A.~Palasyuk, and Y.~Mudryk, Physics-Informed Machine-Learning Prediction of Curie Temperatures and Its Promise for Guiding the Discovery of Functional Magnetic Materials, Chem. Mater. \textbf{35}, 6304 (2023).

\bibitem{Nelson2019}J.~Nelson, and S.~Sanvito, Predicting the Curie temperature of ferromagnets using machine learning, Phys. Rev. Mat. \textbf{3}, 104405(2019).

\bibitem{Curzon1973}A.~E.~Curzon and G.~H.~Chlebek, The observation of face centred cubic Gd, Tb, Dy, Ho, Er and Tm in the form of thin films and their oxidation, J. Phys. F: Metal Phys. \textbf{3}, 1 (1973).

\bibitem{Liu2020}L.~L.~Liu, Y.~Zheng, and W.~Sun, Precipitation of unusual FCC-structured Gd platelets within the thin matrix of a Mg-Gd solid solution, Mater. Lett. \textbf{276}, 128255 (2020).

\bibitem{Teatum1968} E. T. Teatum, K. A. Gschneider Jr., J. T. Waber, Compilation of Calculated Data Useful in Predicting Metallurgical Behavior of the Elements in Binary Alloy Systems, Los Alamos Scientific Laboratory of the University of California, USA, 1968.

\bibitem{Elmers_PRB_13} H. J. Elmers, A. Chernenkaya, K. Medjanik, M. Emmel, G. Jakob, G. Sch\"onhense, D. Gottlob, I. Krug, F. M. F. De Groot, and A. Gloskovskii, Exchange coupling in the correlated electronic states of amorphous GdFe films, Phys. Rev. B {\bf 88}, 174407 (2013).

\bibitem{Chuan_PRB_20} C. W. Chuang, F. M. F. De Groot, Y. F. Liao, Y. Y. Chin, K. D. Tsuei, R. Nirmala, D. Malterre, and A. Chainani, Hard x-ray photoemission spectroscopy of GdNi and HoNi, Phys. Rev. B  {\bf 102}, 165127 (2020).

\bibitem{Nguyen_PRB_22} T. L. Nguyen, Th. Mazet, D. Malterre, H. J. Lin, M. Yoshimura, Y. F. Liao, H. Ishii, N. Hiraoka, Y. C. Tseng, and A. Chainani, Hard x-ray photoemission spectroscopy of the ferrimagnetic series Gd$_6$(Mn$_{1-x}$Fe$_x$)$_{ 23}$, Phys. Rev. B  {\bf 106}, 045144 (2022).

\bibitem{Samanta_NA_20} A. Samanta, S. Das, and S. Jana, Ultra-small intermetallic NiZn nanoparticles: a non-precious metal catalyst for efficient electrocatalysis, Nanoscale Adv. {\bf 2}, 417 (2020).

\bibitem{Feliu_AM_03} S. Feliu and V. Barranco, XPS study of the surface chemistry of conventional hot-dip galvanised pure Zn, galvanneal and Zn--Al alloy coatings on steel, Acta Materialia {\bf 51}, 5413 (2003).

\bibitem{Biesinger_ASS_10} M. C. Biesinger, L. W. M. Lau, A. R. Gerson, and R. St. C. Smart, Resolving surface chemical states in XPS analysis of first row transition metals, oxides and hydroxides: Sc, Ti, V, Cu and Zn, Applied Surface Science {\bf 257}, 887 (2010).

\bibitem{Nayak_PRB_15} J. Nayak, M. Maniraj, A. Gloskovskii, M. Krajčí, S. Sebastian, I. R. Fisher, K. Horn, and S. R. Barman, Bulk electronic structure of Zn-Mg-Y and Zn-Mg-Dy icosahedral quasicrystals, Phys. Rev. B {\bf 91}, 235116 (2015).

\bibitem{Chang_APL_12} Y. H. Chang, C. A. Lin, Y. T. Liu, T. H. Chiang, H. Y. Lin, M. L. Huang, T. D. Lin, T. W. Pi, J. Kwo, and M. Hong, Effective passivation of In$_{0.2}$Ga$_{0.8}$As by HfO$_2$ surpassing Al$_2$O$_3$ via in-situ atomic layer deposition, Applied Physics Letters {\bf 101}, 172104 (2012).

\bibitem{Lee_NC_23} S. W. Lee, M. L. Luna, N. Berdunov, W. Wan, S. Kunze, S. Shaikhutdinov, and B. R. Cuenya, Unraveling surface structures of gallium promoted transition metal catalysts in CO$_2$ hydrogenation, Nat Commun {\bf 14}, 4649 (2023).

\bibitem{Debray1970}D.~K.~Debray, W.~E.~Wallace, and E.~Ryba, Magnetic characteristics of lanthanide zinc (LnZn$_2$) inter-metallic compounds, J. Less-Common Met. \textbf{22}, 19 (1970).

\bibitem{Tsai1979}T.~H.~Tsai and D.~J.~Sellmyer, Magnetic ordering and exchange interactions in the rare-earth gallium compounds RGa$_2$, Phys. Rev. B \textbf{20}, 4577 (1979).

\bibitem{Kouvel1968}J.~S.~Kouvel and J.~B.~Comly, Magnetic equation of state for nickel near its curie point, Phys. Rev. Lett. \textbf{20}, 1237 (1968).

\bibitem{Arrott1967}A.~Arrott and J.~Noakes, Approximate Equation of State for Nickel Near its Critical Temperature, Phys. Rev. Lett. \textbf{19}, 786 (1967).

\bibitem{SI} See the supplementary information for the Arrott plots for the different universality classes and DFT Methodology of utilized to calculate T$_{\rm C}$ of all the samples. 

\bibitem{Liu2018}Y.~Liu and C.~Petrovic, Three-dimensional magnetic critical behavior in CrI$_3$, Phys. Rev. B \textbf{97}, 014420 (2018).

\bibitem{Zhang2012}L.~Zhang, B.~S.~Wang, Y.~P.~Sun, P.~Tong, J.~Y.~Fan, C.~J.~Zhang, L.~Pi, and Y.~H.~Zhang, Critical behavior in the antiperovskite ferromagnet AlCMn$_3$, Phys. Rev. B \textbf{85}, 104419 (2012).

\bibitem{Nehla2019}P.~Nehla, Y.~Kareri, G.~D.~Gupt, J.~Hester, P.~D.~Babu, C.~Ulrich, and R.~S.~Dhaka, Neutron diffraction and magnetic properties of Co$_2$Cr$_{1-x}$Ti$_x$Al Heusler alloys, Phys. Rev. B \textbf{100}, 144444 (2019).

\bibitem{Widom1965}B.~Widom, Equation of state in the neighborhood of the critical point, J. Chem. Phys. \textbf{43}, 3898 (1965).

\bibitem{Kaul1985} S.~N.~Kaul, Static critical phenomena in ferromagnets with quenched disorder, J. Magn. Magn. Mater. \textbf{53}, 5 (1985).

\bibitem{Pramanik2009}A.~K.~Pramanik and A.~Banerjee, Critical behavior at paramagnetic to ferromagnetic phase transition in Pr$_{0.5}$Sr$_{0.5}$MnO$_3$: A bulk magnetization study, Phys. Rev. B \textbf{79}, 214426 (2009).

\bibitem{Fisher1972}M.~E.~Fisher, S.-K.~Ma, and B.~G.~Nickel, Critical exponents for long-range interactions, Phys. Rev. Lett. \textbf{29}, 917 (1972).

\bibitem{Phan_JMMM_07}M. H. Phan and S. C. Yu, Review of magnetocloric effect in manganite materials, J. Magn. Magn. Mater. {\bf 308}, 325 (2007).

\bibitem{Kumar_PRB_24} A. Kumar, P. Singh , A. Doyle, D. L. Schlagel , and Y. Mudryk, Multiple magnetic interactions and large inverse magnetocaloric effect in TbSi and TbSi$_{0.6}$Ge$_{0.4}$, Phys. Rev. B \textbf{109}, 214410 (2024).

\bibitem{Fisher1974}M.~E.~Fisher, The renormalization group in the theory of critical behavior, Rev. Mod. Phys. \textbf{46}, 597 (1974).

\bibitem{Fischer2002}S.~F.~Fischer, S.~N.~Kaul, and H.~Kronmüller, Critical behavior of ferromagnets with quenched disorder, Phys. Rev. B \textbf{65}, 064443 (2002).

\bibitem{Yang2021}X.~Yang, J.~Pan, S.~Liu, M.~Yang, L.~Cao, D.~Chu, and K.~Sun, Critical behavior and anisotropic magnetocaloric effect of the quasi-one-dimensional hexagonal ferromagnet PrCrGe$_3$, Phys. Rev. B \textbf{103}, 104405 (2021).

\bibitem{Zhu2003}L.~Zhu, M.~Garst, A.~Rosch, and Q.~Si, Universally Diverging Grüneisen Parameter and the Magnetocaloric Effect Close to Quantum Critical Points, Phys. Rev. Lett. \textbf{91}, 066404 (2003).

\bibitem{Haldar2010}M.~Haldar, S.~M.~Yusuf, M.~D.~Mukadam, and K.~Sashikala, Magnetocaloric effect and critical behavior near the paramagnetic to ferrimagnetic phase transition temperature in TbCo$_{2-x}$Fe$_x$, Phys. Rev. B \textbf{81}, 174402 (2010).

\bibitem{Chandra2012}S.~Chandra, A.~Biswas, S.~Datta, B.~Ghosh, V.~Siruguri, A.~K.~Raychaudhuri, M.~H.~Phan, and H.~Srikanth, Evidence of a canted magnetic state in self-doped LaMnO$_{3+\delta}$ ($\delta = 0.04$): a magnetocaloric study, J. Phys.: Condens. Matter \textbf{24}, 366004 (2012).

\bibitem{Law2018}J.~Law, V.~Franco, L.~Moreno-Ramirez, A.~Conde, D.~Karpenkov, I.~Radulov, K.~P.~Skokov, and O.~Gutfleisch, A quantitative criterion for determining the order of magnetic phase transitions using the magnetocaloric effect, Nat. Commun. \textbf{9}, 2680 (2018).

\bibitem{Franco2006}V.~Franco, J.~S.~Blázquez, and A.~Conde, Field dependence of the magnetocaloric effect in materials with a second order phase transition: A master curve for the magnetic entropy change, J. Magn. Magn. Mater. \textbf{89}, 222512 (2006).

\bibitem{Biswas2013}A.~Biswas, S.~Chandra, T.~Samanta, B.~Ghosh, S.~Datta, M.~H.~Phan, A.~K.~Raychaudhuri, I.~Das, and H.~Srikanth, Universality in the entropy change for the inverse magnetocaloric effect, Phys. Rev. B \textbf{87}, 134420 (2013). 

\bibitem{Gomes2006} M.~B. Gomes and N.~A. de Oliveira, On the magnetocaloric effect in Gd(Zn$_{1-x}$Cd$_x$), Solid State Commun. \textbf{137}, 431 (2006).

\bibitem{Guo2000}H.~Guo and M.~Enomoto, Influence of Magnetic Fields on $\alpha/\gamma$ Equilibrium in Fe--C(--X) Alloys, Mater. Trans. JIM \textbf{41}, 911 (2000).

\bibitem{Blaha2001}P.~Blaha, K.~Schwarz, G.~Madsen, D.~Kvasnicka, and J.~Luitz, WIEN2k: An Augmented Plane Wave Plus Local Orbitals Program for Calculating Crystal Properties, ISBN 3-9501031-1-2 (2001).

\bibitem{Subhan2020}F.~Subhan and J.~Hong, Magnetic anisotropy and Curie temperature of two-dimensional VI$_3$ monolayer, J. Phys.: Condens. Matter \textbf{32}, 245803 (2020).

\bibitem{Deng2018}Y.~Deng, Y.~Yu, Yijun Song, Yichen Zhang, Jingzhao Wang, Nai Zhou Sun, Zeyuan Yi, Yangfan Wu, Y.~Zheng, S.~Wu, J.~Zhu, J.~Wang, X.~Chen, and Y.~Zhang, Gate-tunable room-temperature ferromagnetism in two-dimensional Fe$_3$GeTe$_2$, Nature \textbf{563}, 94 (2018).

\bibitem{Kresse1994}G.~Kresse and J.~Hafner, Norm-Conserving and Ultrasoft Pseudopotentials for First-Row and Transition Elements, J. Phys.: Condens. Matter \textbf{6}, 8245 (1994).

\bibitem{Bobrow2021}E.~Bobrow, J.~Zhang, and Y.~Li, Ferromagnetic percolation transition in a multiorbital flat band assisted by Hund's coupling, Phys. Rev. B \textbf{104}, 064442 (2021).

\bibitem{Maksymenko2012}M.~Maksymenko, A.~Honecker, R.~Moessner, J. Richter, and O. Derzhko, Flat-band ferromagnetism as a Pauli-correlated percolation problem, Phys. Rev. Lett. \textbf{109}, 096404 (2012).

\bibitem{Priour2004} D.~J. Priour, Jr., E.~H. Hwang, and S.~Das Sarma, Disordered RKKY Lattice Mean Field Theory for Ferromagnetism in Diluted Magnetic Semiconductors, Phys. Rev. Lett. \textbf{92}, 117201 (2004).

\end{thebibliography}
\end{document}